\documentclass[sigconf]{acmart}
\AtBeginDocument{%
  \providecommand\BibTeX{{%
    \normalfont B\kern-0.5em{\scshape i\kern-0.25em b}\kern-0.8em\TeX}}}

\copyrightyear{2024}
\acmYear{2024}
\setcopyright{acmlicensed}\acmConference[UIST '24]{The 37th Annual ACM
Symposium on User Interface Software and Technology}{October 13--16,
2024}{Pittsburgh, PA, USA}
\acmBooktitle{The 37th Annual ACM Symposium on User Interface Software
and Technology (UIST '24), October 13--16, 2024, Pittsburgh, PA, USA}
\acmDOI{10.1145/3654777.3676430}
\acmISBN{979-8-4007-0628-8/24/10}
 \setcopyright{acmlicensed}
 \copyrightyear{2024}
 \acmYear{2024}
 \acmDOI{14151.108}

\usepackage{soul}  
\usepackage{color} 
\usepackage{float}
\usepackage{xspace}
\usepackage{listings}
\usepackage{enumitem}
\usepackage{verbatim}
\usepackage{tabularx} 
\usepackage{booktabs} 
\usepackage{array} 
\usepackage{caption} 
\newcommand{\eg}{{\it e.g.,\ }}

\newcommand{\etc}{{\it etc.}}
\newcommand{\ie}{{\it i.e.,\ }}

\usepackage{booktabs}
\usepackage{array}
\usepackage{multirow}
\definecolor{oxfordblue}{rgb}{0.0, 0.13, 0.28}
\definecolor{harvardcrimson}{rgb}{0.79, 0.0, 0.09}
\definecolor{dartmouthgreen}{rgb}{0.05, 0.5, 0.06}
\definecolor{princetonorange}{rgb}{1.0, 0.56, 0.0}
\definecolor{yaleblue}{rgb}{0.06, 0.3, 0.57}
\definecolor{usccardinal}{rgb}{0.6, 0.0, 0.0}
\definecolor{uclablue}{rgb}{0.33, 0.41, 0.58}
\definecolor{msugreen}{rgb}{0.09, 0.27, 0.23}
\definecolor{cornellred}{rgb}{0.7, 0.11, 0.11}
\definecolor{pomegranate}{RGB}{192, 57, 43}
\definecolor{anti-pomegranate}{RGB}{43,178,192}
\definecolor{alizarin}{RGB}{231, 76, 60}
\definecolor{anti-belize}{RGB}{185, 41, 56}
\definecolor{belize}{RGB}{41, 128, 185}
\definecolor{peter}{RGB}{52, 152, 219}
\definecolor{green}{RGB}{22, 160, 133}
\definecolor{anti-green}{RGB}{160,22,118}
\definecolor{turquoise}{RGB}{26, 188, 156}
\definecolor{pumpkin}{RGB}{211, 84, 0}
\definecolor{anti-pumpkin}{RGB}{0,22,211}
\definecolor{carrot}{RGB}{230, 126, 34}
\definecolor{wisteria}{RGB}{142, 68, 173}
\definecolor{anti-wisteria}{RGB}{99,173,68}
\definecolor{amethyst}{RGB}{155, 89, 182}
\definecolor{nephritis}{RGB}{39, 174, 96}
\definecolor{anti-nephritis}{RGB}{174,39,117}

\newcommand{\pzh}[1]{{\color{black} #1}}
\newcommand{\peng}[1]{{\color{black} #1}}
\newcommand{\zhenhui}[1]{{\color{black} #1}}

\newcommand{\tianjian}[1]{{\color{black} #1}}

\newcommand{\reversion}[1]{{\color{black} #1}}
\newcommand{\name}{{\textit{ComPeer}}}
\newcolumntype{L}{>{\hsize=0.5\hsize}X}
\newcolumntype{R}{>{\hsize=1.2\hsize}X}
\newcolumntype{C}{>{\hsize=1.3\hsize}X}
\begin{document}

\title[\name{}: A Generative Conversational Agent for Proactive Peer Support]{
\name{}: A Generative Conversational Agent \\ for Proactive Peer Support
}
 \author{Tianjian Liu}
 \email{liutj9@mail2.sysu.edu.cn}
 \affiliation{%
   \institution{Sun Yat-sen University}
   \city{Guangzhou}
   \country{China}
 }

\author{Hongzheng Zhao}
 \authornotemark[1]
 \email{zhaohz23@mail2.sysu.edu.cn}
 \affiliation{%
   \institution{Sun Yat-sen University}
   \city{Zhuhai}
   \country{China}
 }
 
 \author{Yuheng Liu}
  \authornote{Both authors contributed equally to this research.}
 \email{liuyh357@mail2.sysu.edu.cn}
 \affiliation{%
   \institution{Sun Yat-sen University}
   \city{Zhuhai}
   \country{China}
 }
 
   \author{Xingbo Wang}
 \email{xiw4011@med.cornell.edu}
 \affiliation{%
   \institution{Cornell University}
   \city{New York}
   \country{USA}
 }
 \author{Zhenhui Peng}
 \authornote{The Corresponding author}
 \email{pengzhh29@mail.sysu.edu.cn}
 \affiliation{%
   \institution{Sun Yat-sen University}
   \city{Zhuhai}
   \country{China}
 }

\begin{abstract}

\tianjian{Conversational Agents (CAs) acting as peer supporters have been widely studied and demonstrated beneficial for people's mental health. 
However, previous peer support CAs either are user-initiated or follow predefined rules to initiate the conversations, which may discourage users to engage and build relationships with the CAs for long-term benefits. 
In this paper, we develop \name{}, a generative CA that can proactively offer adaptive peer support to users. \name{} leverages large language models to detect and reflect significant events in the dialogue, enabling it to strategically plan the timing and content of proactive care.
In addition, \name{} incorporates peer support strategies, conversation history, and its persona into the generative messages.
\zhenhui{Our one-week between-subjects study (N=24) demonstrates \name{}'s strength in providing peer support over time and boosting users' engagement compared to a baseline user-initiated CA.} 
We report users' interaction patterns with \name{} and discuss implications for designing proactive generative agents to promote people's well-being.}

\end{abstract}





\maketitle

\section{Introduction}
\peng{
Peer support refers to the social support exchanged among peers who have similar experience \cite{mead2001peer} and has become an important source for promoting people's well-being \cite{kef2004role,buchanan2008context,drysdale2022feasibility,peng_chi21,peng2020exploring,Li_Wu_Liu_Zhang_Guo_Peng_2024}. 
By sharing advice and feelings with each other, peers build and strengthen their social relationships \cite{o2011healthy}, which could benefit their mental health for a long term \cite{dennis2003peer,faulkner2012helping,thoits2011mechanisms}. 
However, a supportive human peer could not be always available for anyone. 
To mitigate this concern, researchers in Human-Computer Interaction have developed a variety of conversational agents (CAs) acting as peers and demonstrated their usefulness to offer accessible peer support \cite{narain2020promoting,potts2023multilingual,qiu2023psychat,lee2019caring}. 
}

\peng{
However, previous peer support CAs can fall shorts in two aspects. 
First, they are mostly user-initiated, \ie users need to select a topic or send a message to start each round of conversation with the agent. 
This user-initiated design could discourage users to converse with the CAs for a long term and form a close interpersonal relationship \cite{croes2021can}, which is key to the success of peer support \cite{adame2008breaking}. 
In contrast, the proactive messages from CAs can keep the conversation alive and improve conversation productivity \cite{chaves2021should}. 
Literature on peer support also encourages peers to engage in mutual conversations rather than unilateral conversations \cite{sprecher2013effects,watson2017mechanisms}. 
While few CAs can initiate the conversations by asking help from or providing social support to users \cite{lee2019caring,meurisch2020exploring}, the timing and content of their proactive conversations are normally fixed by pre-defined rules. 
The fixed rules can increase the possibility of offering help in a wrong context, which impairs users' perceptions with the CAs \cite{pan2023desirable}. 
Second, previous peer support CAs usually do not learn about users (\eg recent mentally challenging issues and feelings) from the dialogues. 
This limits the CAs' ability to get familiar with users and base the messages on the learned information about users, which can help users feel being understood \cite{croes2023your} and maintain engagement with the CAs \cite{silvervarg2013iterative,chaves2021should}. 
}

\peng{
In this paper, we propose \name{} (\textbf{Com}panion for \textbf{Peer} support), a generative conversational agent that can plan and proactively initiate the peer support chats based on the learned information about users from the \tianjian{conversation history}. \footnote{The code
 is located at \href{https://github.com/liutj9/ComPeer}{https://github.com/liutj9/ComPeer}.} 
Recent advances in large language models (LLMs) have enabled building intelligent agents that can plan, reflect, and proactively take actions when interacting with their environment \cite{park2023generative}. 
However, little is known about the design, effectiveness, and user experience of a LLM-powered CA for proactively offering peer support. 
Questions arise as what are the design principles of the such a CA for proactive peer support, how to enable the CA to adaptively plan the proper timing and content of proactive messages, and how would users with mental health challenges perceive and interact with the proactive CA. 
\pzh{The target users of \name{} are those who are experiencing stressful issues but have not been diagnosed with mental health diseases like depression or eating disorder. 
One representative user group is the university students who are found to frequently encounter mentally challenging issues \cite{robotham2006stress,pascoe2020impact}. 
}
}

\peng{
To begin with, we get insights from literature and work with five experienced users of role-playing chatbots and three students majored in psychology to iterate the design principles of \name{}. 
\tianjian{The principles highlight the importance of the consistent persona as a peer, the proper content and timing of proactive messages, and the usage of psychological strategies in a proactive peer support CA. }
Based on the derived design principles and inspired by \citet{park2023generative}, we adopt GPTs to develop the cognitive architecture of \name{}. 
The \textbf{Memory} module stores all the dialogues between \name{} and the user. 
The \textbf{Schedule} module plans the timings and content that \name{} will proactively send to the user in the near future. 
There are two sources for the schedule. 
First, after one round of conversations (\eg no message from the user within five minutes in our case), the \textbf{Event Detector} extracts the user events (\eg emotional states, plans, challenging issues) from the user's messages of this round and infers the proper timing of each event to proactively talk about it later today. 
Second, at the beginning of a day (\ie 0:00), \name{} invokes the \textbf{Reflection} module to ask itself what it learns about the user from the conversations yesterday, based on which it initializes the schedule on that day. 
The \textbf{Dialogue Generation} module incorporates peer support strategies, 
conversation history, and \name{}'s persona in each generated message. 
}


\peng{
We conduct a one-week between-subjects study with 24 participants to evaluate the effectiveness and user experience of \name{}. 
The baseline user-initiated CA has similar modules with \name{} except the Event Detector, Schedule, and Reflection modules that are responsible for proactive peer support. 
\reversion{The results show that \name{} performs better in making users feel relieved over time \zhenhui{in the one-week study.}}
Participants with \name{} are more satisfied with its proactive messages that care for their states and work than those self-disclosing \name{}'s events.
\reversion{Moreover, \name{}'s proactive messages promote participants' interaction, where they share more their feelings and lives with \zhenhui{\name{}}.}
We offer more qualitative findings to support the \name{} strengths above from an extended one-week study in which all the 24 participants use \name{}. 
We report users' concerns on \name{} and discuss insights from our study for developing proactive agents and proactive generative agents to improve people's mental health. 
}


\peng{
Our contributions are three-fold. 
First, we present a novel generative conversational agent named \name{} that can learn about users from the dialogues and plan the timing and content of proactive messages to offer peer support in the near future. 
Second, our user study demonstrates the usefulness of \name{}'s proactive messages in relieving users' stress, and maintaining user engagement. 
Third, we offer a set of design implications into designing and developing proactive conversational agents with large language models for healthcare scenarios. 
}





\peng{
\textbf{Ethics and Researcher Disclosure}. 
We shape the work by our experience with and observation on people who struggle in stressful issues and desire peer support. 
The authors have experience of exchanging social support with peers like friends, classmates, colleague, and strangers online. 
Two of the authors have experiences and publications that study technologies for supporting people with mental health concerns. 
Our research team obtains IRB approval for broader research projects on patients' and caregivers' practices of healthcare service systems and online communities.
We do not store any personally identifiable information such as participants' names in the design process and evaluation study of \name{}. 
We inform participants that they can request to delete their conversation data with \name{}, and researchers are not allowed to share conversation data to others. 
We keep in touch with the participants every day during the study, inform that they can quit the study at any time if they feel uncomfortable about it, and suggest them to seek professional healthcare services if they face severe mental health problems. 
}

\section{Related Work}
\peng{
\subsection{Peer Support and Conversational Agents}

Peer support is commonly defined as ``social and emotional support, frequently coupled with instrumental support, that is mutually offered or provided by persons having a mental health condition to others sharing a similar mental health condition, to bring about a desired social or personal change'' \cite{mead2001peer}. 
In general, peer support is beneficial for people experiencing 
challenging issues such as depression \cite{pfeiffer2011efficacy} and diabetes\cite{boothroyd2010peers}. 
One key to the success of peer support is the close interpersonal relationship among the peers, which positively predicts the amount of exchanged social support and improvement of psychological health \cite{campos2014familism}. 
According to social penetration theory, deeper and broader self-disclosure among peers can lead to a deeper sense of intimacy \cite{skjuve2022longitudinal,skjuve2021my}. 
Nevertheless, a supportive and closely related human peer is not always available for many people suffering from mental health concerns. 
}





\peng{
Conversational agents (CAs) are promising alternatives to offering peer support to anyone in need \cite{nordberg2019designing,wang2021cass,lee2019caring,qiu2023psychat,liu-etal-2021-towards,jo2024understanding}. 
These CAs commonly incorporate effective psychological skills and strategies in their messages, to provide users with enhanced social support.
Nevertheless, these previous peer support CAs are mostly user-initiated (\eg \cite{liu-etal-2021-towards,qiu2023psychat,wang2021cass}) or follow pre-set rules to initiate the conversation with users at certain timings (\eg \cite{lee2019caring}). 
Such interaction mechanisms could discourage users from maintaining engagement and building a relationship with the CAs \cite{croes2021can, chaves2021should}. 
For one thing, as reported in a longitudinal study in which participants interact with a user-initiated social chatbot over a 3-week period \cite{croes2021can}, participants' level of self-disclosure, the perceived quality of the interactions and the perceived empathy of the chatbot decreased after each interaction, and participants' feelings of friendship with the chatbot were low. 
For another, the pre-set rules could lead to untimely and irrelevant proactive messages of CAs, making the users feel annoyed \cite{portela2017new,liao2016can} and disruptive \cite{chaves2018single,silvervarg2013iterative} 

Our work aligns with previous CAs for peer support by adopting psychological strategies in the conversations with users. 
Different from previous CAs, we seek to build a CA that can proactively conduct peer support conversations that adapt to users' situations and explore whether our proactive mechanism can lead to a close human-CA peer relationship in the long term.  
}



\subsection{Design and Effect of Proactive Agents}



\peng{
Researchers in Human-Computer Interaction have started to design and evaluate the proactive manner of intelligent agents \cite{koppula2015anticipating,zhang2015capability,baraglia2016initiative,chakraborti2015planning,peng2019design,zhang2015human}. 
For example, \citet{peng2019design} defined the service robot's proactivity as ``the anticipatory action that robots initiate to impact themselves and/or others''. 
In a decision-making task, they found that users had sufficient opportunities to express their thoughts and feelings to a robot with medium proactivity, \eg actively confirming user need before providing help rather than actively offering help without user's confirmation or passively waiting for user's requests \cite{peng2019design}. 
In the survey on human-CA interaction design, \citet{chaves2021should} suggested that a proactive CA shares initiative with the user and may manifest proactivity when it initiates exchanges, suggests new topics, provides additional information, or formulates follow-up questions. 
For instance, in the healthcare domain, \citet{fitzpatrick2017delivering} proposed the Woebot that assigns a goal to the user and proactively prompts pre-set motivational messages and reminders at fixed timings to keep the user engaged in the treatment. 
\citet{chaves2021should} summarized several benefits of manifesting a CA's proactivity in the conversations, such as 
providing additional and useful information \cite{medhi2017you,avula2018searchbots,peng2019design,chaves2018single}, inspiring users and keeping the conversation alive \cite{silvervarg2013iterative,chaves2018single,silvervarg2013iterative}, recovering the agent from a failure \cite{portela2017new,silvervarg2013iterative}, improving conversation productivity \cite{jain2018evaluating,avula2018searchbots}, as well as guiding and engaging users \cite{fitzpatrick2017delivering,chaves2018single}. 
However, designing a proactive CA is also challenging, especially regarding the timing and relevance of its proactive messages, which are mostly pre-set by the designers of CAs \cite{portela2017new,liao2016can,chaves2018single}.
For example, \citet{liao2016can} investigated proactivity of a CA in a workspace environment and found that the perceived interruption of the CA's proactivity negatively affects users' perceptions of it. 
Besides, the CA's proactive disclosure may cause users' concerns about privacy \cite{duijvelshoff2017use} and being controlled \cite{tallyn2018ethnobot,toxtli2018understanding}. 
}

\peng{
Following the design of proactivity in these previous agents, 
our \name{} initiates the conversations with users by actively disclosing itself and querying their needs for peer support before offering help.  
Instead of pre-setting rules for triggering proactive behaviors, \name{} can adapt the timing and content of its proactive messages based on the ongoing conversations with users. 
We offer quantitative and qualitative findings regarding the effect of \name{}'s proactive messages on the peer support outcome and process. 
}

\subsection{Generative Agents Powered by Large Language Models}


\peng{
The recent advances in large language models (LLMs) have empowered researchers to build the key components of generative agents \cite{park2023generative,wang2023humanoid, sumers2023cognitive}. 
For example, \citet{park2023generative} presented an LLM-powered architecture that enables generative agents to simulate believable human behavior in an interactive sandbox environment. 
The architecture includes a memory module that stores the agent's perceptions of the environment, retrieves the memories related to its current situation, and passes these memories to an LLM to decide its actions \cite{park2023generative}. 
It also has a reflection module that prompts the LLM to extract high-level insights from its memories to enrich the memory module, as well as a plan module that prompts the LLM to adaptively schedule the agent's activities in the near future \cite{park2023generative}. 
Sumers et al. \cite{sumers2023cognitive} proposed Cognitive Architectures for Language Agents (CoALA) that use the LLM to transform observations into text, choose actions, and manage the agent's internal state via processes such as learning and reasoning. 
The architecture has a decision procedure module that interacts with the LLM, the internal memory, and the external memory \cite{sumers2023cognitive}. 
In each decision cycle with the environment, the agent uses retrieval and reasoning to plan its actions \cite{sumers2023cognitive}. 
}
\peng{
Inspired by these generative agents, we use LLMs to build up \name{}'s architecture, including the Memory, Schedule, Event Detector, Reflection, and Dialogue Generation Modules. 
We contribute a new LLM-powered generative conversational agent for proactive peer support and empirical findings on how users perceive and interact with it in real-world contexts.
}


\section{Design Process and Principles of \name{}}
\peng{
In this paper, we specifically target university students, a representative group that commonly encounters stressful issues \cite{robotham2006stress,pascoe2020impact}. 
The language used by the \name{} and the participants in the design process and user study is Chinese. 
We translate the participants' quotes, the prompts to large language models, and the example dialogues to English when they are presented in this paper. 

\subsection{Design Process}
To design \name{} for proactive peer support, we draw insights from related work and worked with five experienced users (U1 - U5, $M_{age} = 21.6, SD_{age}= 1.85$) of other conversational agents (CAs) and three psychological students (S1 - S3, $M_{age} = 21.6, SD_{age} = 0.47$). 
The users are recruited via an advertising post in \pzh{an online forum about CAs}, and the students are invited from a local university. 
To begin with, we survey CAs in mental health domains (\eg Vincent \cite{lee2019caring}, WoeBot \cite{fitzpatrick2017delivering}) to get familiar with their messages and effective peer support strategies (\eg self-disclosure, reflect of feelings
\cite{liu2021towards,li2023understanding}). We also outline potential contexts for proactive peer support from the literature on proactive AI systems (\eg \cite{chaves2021should}, \cite{peng2019design}, \cite{meurisch2020exploring}). 
Then, we picture \pzh{six} example dialogues in which \name{} initiates a peer support chat with a user in a specific context, \ie \tianjian{Daily Greetings, Emotional Comfort, Life Sharing, Plan Reminder, Encouragement, and Offer Advice}.
In the examples, \name{} proactively provides suggestions and encourages users to review in one evening based on their exam schedule for tomorrow. 
Next, we conduct semi-structured interviews with each participant to query about their experiences with CAs, feelings and comments about the example proactive dialogues, and suggestions and expectations for \name{}. 

Based on their feedback, we develop the first workable prototype of \name{} with an architecture similar to the final one described in \autoref{sec:architecture}. \reversion{Then, five authors from our team conduct a one-week pilot study using the prototype CA and iterate the designs, \eg adding the long-term memory. After the iteration \zhenhui{within the research team}, we invite \zhenhui{all} participants \zhenhui{(U1 - U5, S1 - S3) mentioned above} to experience the \name{} prototype and iteratively \zhenhui{refine} it based on their suggestions over the following two months.} 
The key refinements \zhenhui{in this process} lie in the \pzh{frequency of proactive messages, length of \name{}'s messages, the prompt format of \name{} personas, and the mechanism of event detection.}
We compile the following design principles for the refined \name{} that is evaluated in our user study (\autoref{sec:user_study}).
}

\subsection{Design Principles}
\peng{
\textbf{DP1. \name{} should have a consistent persona as a peer with backgrounds similar to the user's and reveal it in its messages during the peer support chats.} 
To gain the trust and interest of users, a CA needs to show consistent and vivid personalities \cite{mou2017media,shum2018eliza}. 
Otherwise, as mentioned by three participants, \textit{``the CA would be lifeless and boring during the interaction''} (U1). 
Apart from the consistency, we have the following two sub-principles. 

}

\peng{\reversion{
\emph{\textbf{DP1.1}. The persona should have customizable backgrounds or skills similar to the user's.} 
Peer support requires the provider and the seeker to have similar experiences to achieve resonance and understanding in the conversation \cite{o2017design}.  
Therefore, as our participants are university students, we design the default persona of \name{} prototype as a student who majors in computer science. Moreover, we allow participants to customize the personas based on their background and expectation before the interaction.
During the interaction with the prototype, four participants agree that the personas similar to themselves can make the CA more realistic and reliable.} 
}

\peng{
\textit{\textbf{DP1.2}. The language style of \name{}'s messages should adapt to its persona and match the characteristics of online peer support chats.}
\name{} interacts with users in instant messaging apps via textual messages, which is the only medium for it to offer peer support. 
Two participants strongly suggest that \name{}'s messages, especially the proactive ones, should adapt to its persona, as they view the proactive messages from previous CAs were ``\textit{very rigid}'' (U2). 
Besides, the style of \name{}'s messages should be in line with that used in normal online chats between peers \cite{mou2017media}. 
\reversion{Initially, the average length of \name{} prototype's \zhenhui{messages is} 140.6 (SD = 54.78) Chinese words, which \zhenhui{are considered quite} long and complex by four participants}, while \textit{``online conversations are often brief and relaxed''} (U4). 
They suggest that the length of the messages should be limited, \eg \textit{``between 20 and 50 words''} (U5). 
Two participants further give suggestions on the composition of \name{}'s messages. 
``\textit{It shouldn't end every message with encouragement, as a real friend wouldn't do that in online chats}'' (U3).
}

\peng{
\textbf{DP2. The content of the \name{}'s proactive messages should include sharings about its daily life and/or concerns on the user's specific conditions.} 
Many existing CAs (\eg \cite{silvervarg2013iterative}, \cite{fitzpatrick2017delivering}) can initiate conversations with users, but the content of their proactive messages is usually limited to greetings and requests for interaction. 
Participants actively share their expectations on the content of \name{}'s proactive messages as summarized below. 
}

\peng{
\emph{\textbf{DP2.1}. The \name{}'s proactive messages should disclose itself based on its persona and schedule.} 
Mutual self-disclosure can promote peers' relationship \cite{skjuve2021my} and benefit their perception of receiving peer support \cite{skjuve2021my}. 
Four participants hope that \name{} could share its experience and thoughts.
``\textit{It should share its life and emotions like a friend, making it feel alive}'' (U1).
For example, in the interactions with \name{} prototype, U2 favors its proactive sharing that ``\textit{I'm happy for the basketball game with classmates this afternoon}'', which \tianjian{``\textit{makes me feel it view me as a friend}'' (U2).}  
}

\peng{
\emph{\textbf{DP2.2}. The \name{}'s proactive messages should express concerns on the user's specific conditions.} 
Showing concerns on the user's condition can reflect that \name{} remembers what happened to the user and keep the message focused, which can make the user feel being cared and supported \cite{croes2023your}. 
Two participants explain why the \name{}'s proactive care is essential for them. 
\textit{``\pzh{I can know that the agent is specifically caring for me, while in real life, I seldom receive proactive support from other people}''} (U4). 
}

\tianjian{\textbf{DP3. The frequency of the \name{}'s proactive peer support should adapt to the content of its intended proactive message and the user's reaction to the previous proactive message}. \reversion{
\zhenhui{Initially,} the \name{} prototype is like a fixed-rule CA when sending proactive messages (\eg providing care every two hours), which is deemed inflexible and dull by five users in the design process. ``\emph{It just like a timer that sends pre-set messages to me}'' (U1). Thus, we iterate the event detector and the schedule module to improve proactive messaging. However, it still automatically sends proactive messages whenever the timing of the events occurred, averaging 6.6 (SD = 1.88) times per day. Our participants complain that these proactive messages were too frequent, which would be disturbing especially when they see a new proactive message before replying to a previous one. ``\emph{It is time-consuming and exhausting to check and reply to so many messages in a day}'' (U2). ``\textit{I delay responding to some non-urgent messages from the agent, but it eagerly sends another half an hour later, making me a bit annoyed}'' (U4).
Therefore, before sending proactive messages,
\name{} should infer if the content of the message would be important for the user and control the amount of proactive messages within a day. S2 further suggestes that ``\textit{If the user doesn't respond to a proactive message within a time, it should pause the next scheduled one.}'' }}

\peng{
\textbf{DP4. \name{} should apply effective peer support strategies to help users relieve their stresses.} 
Enacting effective psychological strategies in the conversations can improve the efficacy of peer support chats \cite{liu2021towards,fitzpatrick2017delivering,qiu2023psychat}. 
In an early test of \name{}, two users discover the conversations with \name{} are inefficient and helpless due to \name{}'s lack of support strategies, ``\emph{I shared my exam failure, but it only inquired about the content instead of offering comfort}'' (U1). 
Our psychology participants help us identify the proper peer support strategies that \name{} can use, such as \pzh{Affirmation and Reasurance, Self-disclosure, and Reflection of feelings \cite{liu2021towards,li2023understanding}}, which can provide users with more effective support. 
}

\section{Design and implementation of \name{}}
\label{sec:architecture}

\begin{figure*}
    \centering
    \includegraphics[width=1.1\linewidth]{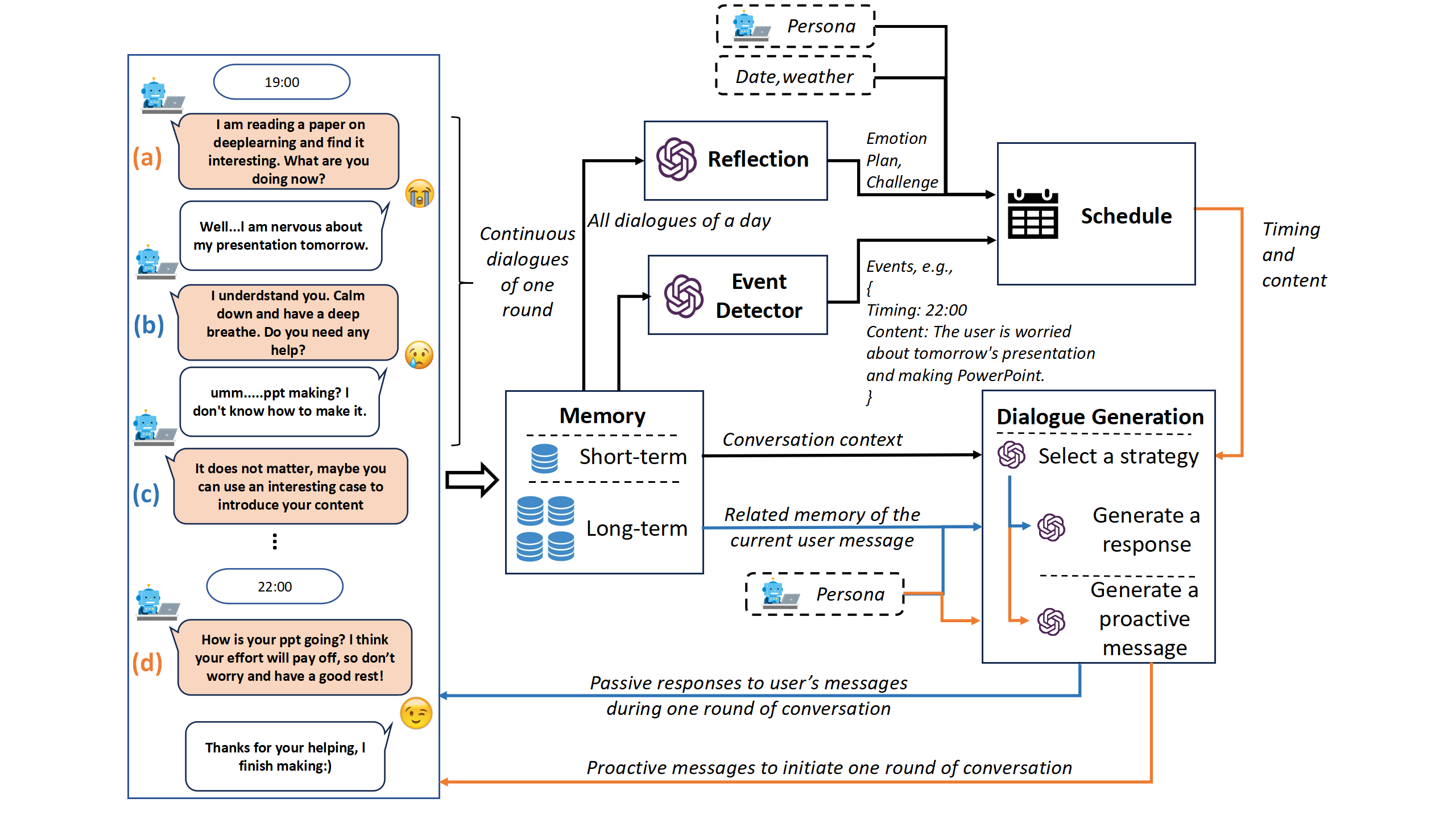}
    \captionsetup{justification=centering}
    \caption{Architecture and an example user scenario of \name{}, a generative agent for proactive peer support.}
    \label{fig:workflow}
\end{figure*}

\peng{
In this section, we present the design and implementation of \name{} based on the design principles derived from the design process. 
We first present a user scenario of \name{} to illustrate how each of its modules works (\autoref{fig:workflow}) and then detail the implementation of each module in the \name{}'s architecture. 
}

\peng{
\subsection{User Scenario and Architecture of \name{}}
}

\peng{
In this scenario (\autoref{fig:workflow}), we describe how \name{} had a peer support chat with Carl, a junior student who majored in computer science.  
Carl initialized \name{}'s persona (\autoref{sec:persona_initialization}) as \tianjian{a student who has the same major and hobbies as him.}
\tianjian{After four days of interactions}, 
one evening at 19:00, \name{} proactively sent a message to Carl, ``\emph{I am reading a paper on deep learning and find it interesting. What are you doing now?}'', 
Carl was anxious about his upcoming presentation on the next day and expressed his anxiety, ``\textit{Well... I am nervous about my presentation tomorrow.}''. 
\name{} comforted him using the ``Affirmation and Reassurance'' strategy, ``\textit{I understand
you. Calm down and have a deep breath. Do you need any help?}''.
They then discussed how to prepare a PowerPoint for the presentation. 
Based on the content of the conversation, \name{} proactively checked in with Carl to see how he was doing with his PowerPoint at 22:00 and encouraged him, \textit{``How is your ppt going? I think your effort will pay off,
so don't worry and have a good rest!''}. 
Carl expressed gratefulness to \name{}'s proactive care and felt being supported. 
}



\peng{
To enable \name{} to provide such proactive peer support, we build up an architecture (\autoref{fig:workflow}) of generative conversational agents powered by large language models (LLMs). 
The Memory module (\autoref{sec:memory_module}) stores all \tianjian{conversation history} between the user and \name{}. 
From each round of stored conversations (\eg no message from the user within five minutes in our case), the event detector module (\autoref{sec:event_detector}) extracts a user event (\eg ``The user is worried about tomorrow's presentation and making PowerPoint'') and infers a proper moment later that day (\eg 22:00 tonight) for \name{} to proactively talk about it in the near future. \tianjian{These detected events with inferred timings of proactive messaging are sent to the schedule module, which decides whether each message will be proactively sent to the user.}
At the beginning of each day, the reflection module (\autoref{sec:reflection_module}) triggers \name{} to reflect on the user's conditions, emotions, challenges, and plans based on the conversations yesterday. 
The output reflections will help the schedule module initialize the planned proactive messages on that day. 
Lastly, the Dialogue Generation module (\autoref{sec:dialogue_generation_module}) takes into considerations the \name{}'s persona, \tianjian{relevant memory} 
and proper peer support strategies to generate a message to initiate the conversation or respond to the user. 
}

\subsection{Personas Initialization}
\label{sec:persona_initialization}
\peng{
Following the persona settings used in previous works \cite{park2023generative,shum2018eliza}, we describe the \name{}'s persona with its name, age, gender, personality, occupation or major, background, hobbies, and language style. 
In addition, the persona describes its relationship with the user, which could enhance its understanding of the user during the conversation. 
For instance, the \name{}'s persona in the user scenario (\autoref{fig:workflow}) is an outgoing and gentle student (\textit{personality}) who majors in artificial intelligence (\textit{major}) and has previously studied geographical sciences (\textit{background}). He has interests in cryptography, basketball, and video games (\textit{hobbies}). He prefers using exclamatory sentences in conversations (\textit{language style}) and is a classmate of the user (\textit{relationship with the user}).










At the beginning of our user study, participants can customize the persona of \name{} based on their expectations of a supportive peer (DP1.1).  
When saving the persona as part of the prompt to LLMs, we add four pairs of example dialogues between \name{} and the user, which enable the LLMs to conduct in-context learning for generating consistent peer-like messages \cite{li2023chatharuhi} (DP1.2). \reversion{Besides, we prompt \name{} to be compassionate in the dialogue generation module regardless of its persona.} The prompt template of \name{}'s personas is listed in \autoref{sec:prompt of personas} in Appendix. 
}

\peng{
\subsection{Memory Module}
\label{sec:memory_module}
}

\peng{
We refer to previous work on generative agents \cite{park2023generative, Significant_Gravitas_AutoGPT} to design a memory module, which 
\pzh{maintains a streaming record of the conversations between \name{} and the user.
It is a list of memory objects, where each
object contains one pair of \name{}-user messages, \tianjian{\eg [\{"role": "user", "content": "I feel nervous because the deadline of homework is coming."\}, \{"role": "Assistant", "content": "I understand you, try your best to finish it! I am there to help you!"\}]}. 
}
When \name{} initiates a chat or responds to the user, it needs to refer to the conversation history to keep its messages consistent and logical. 
In our practice, we get inspired from human's short-term and long-term memories \cite{atkinson1968human} to prepare two databases for storing the agent's memories. 
One database maintains the context during the conversation, 
which serves as the short-term memory for generating \name{}'s messages \cite{zhang2023memory}.  
In our case, the context can contain at most 20 messages \footnote{We experiment different numbers of previous messages as the context. We find out that using at most 20 messages as the context can balance the trade-offs between costs of prompting LLMs and quality of the generated message in our case.} of either \name{} or the user prior to \name{}'s intended message.
The other database stores the long-term memory  \cite{cowan2008differences}, \ie all the memory objects \pzh{except those in the short-term database}. 
We also encode each object of the long-term memory into a vector using \tianjian{OpenAI's text-embedding-ada-002 model \cite{ke2024development}, 
which helps \name{} retrieve related memory to the current context in the later dialogue generation module. 
}
}



\peng{
\subsection{Event Detector Module}
\label{sec:event_detector}
}

\peng{
While users can actively seek for comfort and support from \name{}, the most unique design of our \name{} is its ability to initiate a peer support chat with appropriate content at a proper timing (DP2, DP3). 
The event detector module is responsible for preparing the content and inferring the timing for the later schedule module to determine \name{}'s proactive messages. 
Specifically, in the prompt to the GPT-4, we define an event as a specific incident or occurrence that brings about \tianjian{stress} and has a significant impact on the individual \cite{tennant2002life}. 
\pzh{We consider an appropriate event for \name{}'s proactive care as \tianjian{poor physical condition \cite{schwarzer1991social}, negative feeling \cite{thoits1985social}, challenges \cite{schwarzer2007functional}, and plan of the user\cite{deci2000and}, \peng{which would desire social support from others}.}} 
\pzh{We constrain the inferred timing of the event as a time point later that day, as our trials on the prompts indicate that the GPTs often fail to infer a proper timing in the subsequent days.
} 
An example event is shown in \autoref{fig:workflow}. 




}


\peng{
The event detector extracts an event from each round of conversation after it ends, \pzh{\ie \name{} does not receive a message from the user within 5 minutes}, \tianjian{which can avoid detecting redundant conversation in a round of conversation.} 
The event detector queries GPT-4, a state-of-the-art LLM at the time of \name{}'s development, using the chain-of-thoughts (COT) prompts, which can enhance the LLM's ability of inference given a few data samples \cite{chen2023unleashing}. 
We experiment various prompts in the development process and iteratively improve the prompt to achieve acceptable performance in event detection. 
\pzh{Lastly, we prepare \tianjian{five} 
\tianjian{examples} about \tianjian{the detection process of physical condition, negative feeling, challenges, plans, and no event} in the prompt.}
The example COT prompts used in the event detector are shown in \autoref{sec:prompt of event_detector} in Appendix.

}

\peng{
\subsection{Reflection Module}
\label{sec:reflection_module}
}
\peng{
While the events detected from one round of conversation help the schedule module update the planned proactive messages on that day, \name{} needs to gain a higher-level understanding about the user to initialize a consistent plan of offering proactive care on the next day. 
Consider a case in which the user had several rounds of conversations with \name{} on a day. 
\name{} could have proactively cared for the user on the detected events about anxiety on the presentation and a plan for going shopping tomorrow. 
We hope that \name{} can proactively initiate conversations on the next day based on what it learns from the user on that day, as it would be beneficial to care for the user again on the event that causes the user's anxiety or could make the user happy. 
}

\peng{
Inspired by \cite{park2023generative}, at the beginning of each day, the reflection module asks three questions, \ie \emph{``Does the user feel negative emotions due to something?}'', ``\emph{Are users facing challenges or difficulties?}'', and ``\emph{What plan does the user has in tomorrow?}'', on all the dialogues of the previous day in the prompt to GPT-4. 
The module outputs three reflected thoughts, each for one question, \eg 
``1) \emph{The user is nervous about presentation}, 
2) \emph{There is an presentation that user has to deliver}, 3) \emph{The user will make presentation tomorrow''.} 
These thoughts will be sent to the schedule module as a condition for generating the initial schedule of a day. 
}


\begin{table*}
\centering
\caption{\peng{Descriptions and examples of candidate peer support strategies used by \name{}. \name{} can select anyone of all the strategies for generating a passive response to users message. However, \name{} is allowed to use \emph{self-disclosure}, \emph{inquiring}, \emph{affirmation and reassurance}, \emph{invite users to think} for generating a proactive message, as the usage of other strategies needs the user starts the conversation first.}}
\label{tab:strategies}
\begin{tabularx}{0.9\textwidth}{L C R} 
\hline
\textbf{Strategy} & \textbf{Description} & \textbf{Example} \\
\hline
Self-disclosure & Disclose personal information to users, including but not limited to the counselor’s own similar experiences, feelings, behaviors, thoughts, \etc. & "I also have a similar experience! I did such a thing last time!" \\
\hline
Inquiring & Explore users' subjective experiences or ask users to concretize the imprecise factual information & "Are you feeling better now? How do you feel now?" \\
\hline
Affirmation and Reassurance & Affirm users' strengths, motivations, and abilities, and normalize users' emotions and motivations, and provide comfort, encouragement, and reinforcement. & "I believe you can! It will get better soon! I will support you no matter what!" \\
\hline
Invite users to think & Answer the questions that users ask about the conversation topics. & "Taking a deep breath may be a good way to relax, but it's important to first identify the root cause of the problem." \\
\hline
Reflection of feelings & Use tentative or affirmative sentence patterns to explicitly reflect the users' mood, feelings, or emotional states. & "I understand your current annoyance." \\
\hline
Restatement or Paraphrasing & Reflect the content and meaning expressed in users' statements to obtain explicit or implicit feedback from users. & "It sounds like you think everyone is ignoring you, right?" \\
\hline
Answer & Answer the questions that users ask about the conversation topics. & "I think you need to read the book 'Learning Neural Networks' now." \\
\hline
\end{tabularx}
\end{table*}

\peng{
\subsection{Schedule Module}
\label{sec:schedule_module}

\peng{The schedule module handles the initialization of the plan for proactive care on a day based on the input from the reflection module, the update of the planned proactive messages on that day based on the input from the event detector, and adjustment of the frequency of proactive care based on the content of each proactive message and the user's reaction to this message (DP3).}
}
\peng{
\subsubsection{Initialization of the schedule on a day}
At the beginning of each day, \name{} will initialize schedules for the day. 
Apart from the reflected thoughts about the user on the previous day, the schedule module incorporates the real-world information (\eg date, weather) of that day into the prompt for generating a schedule closely related to the user's real life.
For example, if the reflection module indicates that the user had a plan or encountered a challenge on previous day, \name{} could remind the user of the plan or care for the user about the challenge on that day. 
If the weather is rainy, \name{}'s schedule had better not include outdoor activities. 
One example of the real-world information in the prompt is: ``Today is Friday. The weather of today is rainy. The lowest temperature today is 4°C and the highest temperature is 9°C''. 
Besides, the prompt for schedule initialization also consists of \name{}'s persona to guide the GPT-4 to generate the schedule with events that a peer should care (DP1.1). 
By considering the reflected thoughts about the user, real-world information, and \name{}'s persona in the prompt, the schedule module can initialize a schedule on that day. The specific schedule template is shown in \autoref{sec:schedule_generation} in Appendix.}















\peng{
\subsubsection{Update of the schedule and adjustment of the frequency of proactive messaging on that day}
Once the event detector extracts an event from one round of conversation, the schedule module inserts the extracted event in the priority queue of the schedule. 
\reversion{If the conversation do not contain the event, the detector will output "" (no event). In this case, \name{} relies on the initial schedule to plan for proactive messages.} As a default setting, when the planned timing of an event comes, the schedule module will dequeue the event and send it to the dialogue generation module to initiate a conversation. 
However, following the default setting, if the planned events in the schedule are too much, the frequent proactive messages would be annoying. 
To adjust the frequency of proactive messaging, the schedule module evaluates the content of the planned event and the user's reaction to \name{} previous proactive message (DP3). 
As suggested by \cite{silvervarg2013iterative}, a proactive message related to the user's information will help the user feel be cared and supported and maintain higher engagement. 
\tianjian{Therefore, we allow a LLM to infer the importance value (a score between 0 and 1) of each planned event.}  
\pzh{For example, an event like having a dinner with friends or reading books in the library could be less important than an event that is more about the user's mental state, \eg the user is depressed for the failure of an exam.} 
To encourage proactive care on more ``important'' events while ensuring that all events have a chance to be proactively sent, we design a randomized mechanism for the schedule module to make decisions. 
When the timing of an event arrives, the module generates a random number between 0 and 1. 
If the event has an importance value larger than the random number, it will be sent to the dialogue generation module. 
Besides, if the user does not reply to \name{}'s previous proactive message for a certain period, \eg three hours in our study, the schedule module will not send other planned events within this period to the dialogue generation module. 
}

\peng{
\subsection{Dialogue Generation Module}
\label{sec:dialogue_generation_module}
}
\peng{The dialogue generation module is responsible for generating peer-like, consistent, and supportive messages to respond to or initiate a conversation with the user (DP1, DP4). }
To achieve this purpose, we prompt GPT-4 to act as three roles (noted as LLM-1, LLM-2, and LLM-3), which 
\tianjian{can simplify the generation tasks by allowing the LLM focus on one task at a time \cite{talebirad2023multi}.} 
\peng{LLM-1 takes in \peng{the conversation \tianjian{history}
stored in the} short-term memory and selects \pzh{one \tianjian{or more}} peer support strategy \peng{that can be used in \name{}'s message.} 
As suggested by our psychological students in the design process, the candidate strategies for generating passive responses are self-disclosure \cite{lee2020hear,lee2020designing}, inquiring \cite{li2023understanding}, affirmation and reassurance \cite{liu2021towards,cheng2022improving}, invite users to think \cite{li2023understanding,qiu2023psychat}, reflections of feelings \cite{cheng2022improving,highlen1977effect}, restatement or paraphrasing \cite{lee2021restatement} and answer \cite{li2023understanding}. 
\autoref{tab:strategies} shows the description and example of each strategy. 
As for the generation of proactive messages, the candidate strategies are self-disclosure (DP2.1), inquiring (DP2.2), affirmation and reassurance (DP2.2), and invite users to think (DP2.2). 
The other three strategies, \ie reflection of feelings, restatement or paraphrasing, and answer need user to express, need the user to express feelings first and are not suitable for \name{}'s proactive messaging. 

    










}

LLM-2 generates a passive responding message conditioned on the peer strategy selected by LLM-1, \pzh{the conversation history 
}, \name{}'s persona, and the related long-term memory of the current user message. 
To retrieve related long-term memory, we encode the current user message into a vector using \pzh{\tianjian{text-embedding-ada-002} 
and calculate its cosine similarities with the vectors of memory objects stored in the long-term memory database. 
We select at most three memory objects that have the highest similarity scores for the prompt to LLM-2. 

}







\peng{
To generate proactive messages, we choose to design another prompt to require the LLM (note as LLM-3) to act as a role different to that of LLM-2, because we find in our practice that an LLM has difficulty distinguishing its roles if we prompt it to handle the generations of passive responses and proactive messages at the same time. 
The prompt to LLM-3 considers \name{}'s persona, the event from the schedule module, and the peer support strategy selected by LLM-1. \tianjian{The prompts for LLM-1, LLM-2, and LLM-3 are presented in \autoref{sec:prompt of LLM-1}, \autoref{sec:prompt of LLM-2}, and \autoref{sec:prompt of LLM-3} in the Appendix.}}

\section{User study}

\label{sec:user_study}
\peng{
To evaluate the effectiveness and user experience of \name{}'s proactive peer support, we conduct a one-week between-subjects user study with 24 participants compared to a baseline conversational agent (CA) that can not proactively initiate the conversation with users. 
After the one-week study, we invite all participants, either in the \name{} or baseline group, to use \name{} for another week, which provide us with more qualitative findings especially on the long-term user experience of \name{}.
Our research questions (RQs) in the user study are: 

\textbf{RQ1:} \pzh{How would interacting with \name{} affect the users' management of their stress?}

\textbf{RQ2:} \pzh{How would users perceive a) the social support provided by \name{}, b) the persona of \name{}, and c) their relationship with \name{}?}

\textbf{RQ3:} \pzh{How would users perceive \name{}'s proactive messages?}

\textbf{RQ4:} \pzh{How would users interact with \name{}?} 

}

\peng{
\subsection{Participants}
We recruit 24 participants \tianjian{(15 males, 9 females, aged 18-25)} via recruitment posts in three universities. 
The inclusion criteria in the background survey are that participants have recently encountered stressful issues and they have interests in interacting with a peer support CA every day in a two-weeks period. 
On average, their \pzh{perceived level of stress measured using PSS-10 \cite{cohen1983global} was $23.63$ ($SD = 3.98$), indicating that most of them perceived high level (score > 19) of stress.}
Their majors vary across computer science, psychology, international politics, electronic engineering, and so on.  \reversion{Their stress primarily originates from academic workload and examinations (16 participants), interpersonal relationships (5 participants), and personal future development (6 participants).
Although only 
two participants report experience of disclosing themselves to a CA or using AI companion applications (\eg Character AI, Replika), 
eighteen participants have experience of using general CAs (\eg ChatGPT, Qwen).}
All participants are native speakers of Chinese, the language used by \name{} in the design and development process. 
\reversion{We randomly divide our participants into the \name{} group (PC1-PC12, 7 males, 5 females, mean age: 20.63, mean PSS-10 score: 24.45) and the baseline group (PB1-PB12, 8 males, 4 females, mean age: 20.08, mean PSS-10 score: 23.54).}

}

\peng{
\subsection{Baseline Conversational Agent}
The primary goal of our one-week between-subject study is to evaluate the effects and user experience of the proactive peer support mechanism we built for \name{}.  
Therefore, the baseline CA does not have the Event Detector, Schedule, and Reflection modules that are responsible for proactive peer support. 
Nevertheless, it has the same Memory and Dialogue Generation modules as those of \name{} (\autoref{fig:workflow}), which enable it to react to user-initiated messages in a similar way of \name{}. 
Such a CA can not only provide a strong and fair baseline for evaluating the impact of \name{}'s proactive peer support but also offer more qualitative data for evaluating the usability of the memory and dialogue generation modules in our architecture of generative CAs. 
}

\begin{figure*}
    \centering \includegraphics[width=0.9\linewidth]{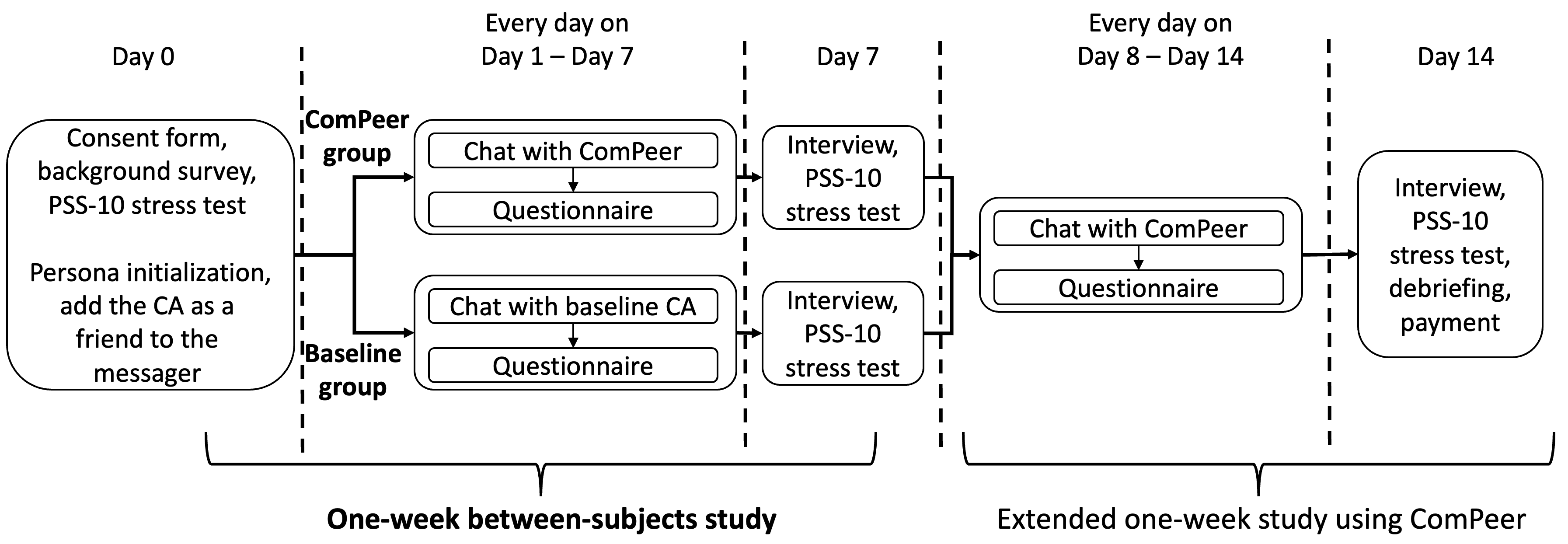}
    \caption{\peng{Procedure of our one-week between-subjects study (\name{} vs baseline conversational agent (CA)) and extended one-week study in which all participants converse with \name{}.}}
    \label{fig:procedure}
\end{figure*}
\peng{

\subsection{Measurements}
\label{sec:measurement}
\peng{
\autoref{tab:scale} summarizes the items used in the questionnaire after participants' interactions with either \name{} or the baseline CA everyday. 
All items are measured on the 7-point Likert scale, with 1 for strongly disagree and 7 for strongly agree.

\textbf{RQ1. Effects on stress management.} 
We use the Pressure Perceive Scale 10 (PSS-10, \pzh{\tianjian{10} items, total scores: 0 - \tianjian{40}}) \cite{cohen1983global} to measure each participant's perceived level of stress at the beginning of the user study, after one week of interactions with the CA, and after two-week of interactions. 
Following \cite{o2018suddenly}, we evaluate the change of participants' perceived stress before and after interacting with our \name{} or the baseline CA. 
Besides, in the questionnaire sent to participants every evening during the study, we ask participants to rate their agreement on the following statement adapted from \cite{mcconchie2022little}: ``\textit{I feel relieved when I talk to the CA.}''
}

\peng{\textbf{RQ2. Perceived Social Support, Persona, and Relationship.} 
To measure how the user perceive the social support provided by \name{}, we adapt five items from \cite{mcconchie2022little, stiles1980measurement}. 
They are ``\textit{The CA is accompanying me}'', ``\textit{The CA is caring about me}'', ``\textit{The CA spares no effort to help me out of difficulties}'', ``\textit{The CA's advice is very important to me}'', and ``\textit{The CA is encouraging me to be a better person in life}''.
As for the perceived persona of the CA, we apply two items from \cite{salminen2018persona}: ``\textit{The CA's response is logical and coherent}'', and ``\textit{The persona of the CA is interesting enough}''. 
We capture how participants perceive their relationship with the CA using three items adapted from \cite{mcconchie2022little}: ``\textit{Chatting with the CA has deepened my relationship with it}'', ``\textit{When I feel in trouble, I am willing to seek help from the CA}'', and ``\textit{When I encounter something joyful, I am willing to share it with the CA}''.
}

\peng{\textbf{RQ3. Perceptions of \name{}'s Proactive Messages.} 
In the questionnaire sent to participants in the \name{} group every evening during the study, we ask them to rate their satisfaction with each proactive message from \name{}, \textit{``I am satisfied with this proactive message''}. 
We also conduct a thematic analysis on the topic of each proactive message from \name{}. 
\pzh{Three authors of this paper first independently label fifty
proactive messages. They then meet and discuss their labels and coding schemes and reach an agreement on five topics, \ie share the CA's daily routine, share the CA's work or study, share the CA's leisure and entertainment, care for the user's work or study, care for user's state. 
They then independently code all the proactive messages (intraclass correlation coefficient $ICC = 0.851$) and resolve disagreement via discussion. }  
We evaluate participants' satisfaction with each topic of proactive messages. 
}



\peng{
\textbf{RQ4. Interaction with and Concerns on the CA.} 
We analyze timing and content of each interaction between the CA and the user. 
Specifically, we calculate the numbers of peer support strategies that \name{} and the baseline CA selected to generate their messages. 
We also conduct a thematic analysis on the topic of each round of conversation from the user's perspective. 
\pzh{Two authors of this paper first independently label fifty 
rounds of conversations. They then meet and discuss their labels and coding schemes and reach an agreement on six topics, \ie concern, want, thoughts or feelings, share happiness, share life, and others. 
They then independently code all the proactive messages ($ICC = 
0.764, 0.807, 0.723, 0.724, 0.772, 0.785$) and resolve disagreement via discussion. } 
}

\subsection{Task and Procedure}
\autoref{fig:procedure} illustrates the procedure of our user study conducted remotely. 
After obtaining participants' consent, background information, and perceived level of stress on Day 0, we guide them to initialize the CA's persona with expected attributes on the CA (\autoref{sec:persona_initialization}) and their personal information (\ie age, major, stress issue) via an online document. \reversion{In the personas initialization, four participants set it with majors or hobbies different from their own, \eg Computer Science vs. Law. }
We help them add the CA as a friend on Tecent QQ, a popular instant messaging app in China. 
In the consent form, we have informed participants that experimenters will look at their conversation record with the CA and that third party technology providers the CA relies on, \ie Tencent, have access to the information. 
We add that participants' personally identifiable information will not be shared for publication purposes and that their voluntary participation means that they can drop out of the study at any point. 
We also state that the CA is not a therapist and that they should seek professional help if any psychological issues are experienced, though we target a non-clinical population.

Every day from Day 1 to Day 7, 
they are asked to have at least one conversation with either \name{} or the baseline CA based on their assigned groups. 
The experimenter send out a questionnaire (\autoref{tab:scale}) to participants every evening and requires them to fill it based on their conversation with the CA on that day. 
On the evening of Day 7, we contact participants, ask them to rate their perceived level of stress, and conduct semi-structured interviews on \tianjian{impact of \name{}, their perceptions of \name{}'s messages, as well as the frequency and content of their interactions with \name{}.}
After the one-week between-subjects study, we enable the proactive peer support mechanism of the baseline CA (\ie it is \name{} now) and invite all participants to chat with \name{} every day for another week. 
Lastly, on Day 14, we conduct another semi-structured interview with each participant, focusing on \tianjian{their feelings towards \name{}, and their perceptions of \name{}'s proactive messages.}
We estimate that each participant spent about 100-150 minutes in our study, including filling questionnaires, interviews, and chatting with the CA. 
After the study, each participant receives about \$10 compensation following the local payment policy.  
}

\begin{table*}[]
\caption{
\peng{Measured items for RQ1-3 in the questionnaire sent to each participant every evening. Participants rate their levels of agreements on the statement in each item on the 7-point Likert scale; 1 - strongly disagree, 7 - strongly agree. CA stands for Conversational Agent.}
}
\label{tab:scale}
\begin{tabular}{llll}
\hline
Research Question                                                                           & ID  & Item                                                                     & Reference                                                                             \\ \hline
(RQ1) Stress Management                                                                     & Q1  & I feel \textbf{relieved} when I talk to the CA.                                   &  \cite{mcconchie2022little}                                           \\ \hline
\multirow{5}{*}{\begin{tabular}[c]{@{}l@{}}(RQ2a) Perceived \\ Social Support\end{tabular}} & Q2  & The CA is \textbf{accompany}ing me.                                               &  \cite{mcconchie2022little,stiles1980measurement}                     \\
                                                                                            & Q3  & The CA is \textbf{caring} about me.                                               &  \cite{mcconchie2022little}                                           \\
                                                                                            & Q4  & The CA spares no effort to \textbf{help} me out of difficulties.                  &  \cite{mcconchie2022little}                                           \\
                                                                                            & Q5  & The CA's \textbf{advice} is very important to me.                                 &  \cite{mcconchie2022little,stiles1980measurement,salminen2018persona} \\
                                                                                            & Q6  & The CA is \textbf{encouraging} me to be a better person in life.                  &  \cite{mcconchie2022little}                                           \\ \hline
\multirow{2}{*}{\begin{tabular}[c]{@{}l@{}}(RQ2b) Perceived \\ Personas\end{tabular}}       & Q7  & The CA's response is logical and \textbf{coherent}.                               &  \cite{stiles1980measurement,salminen2018persona}                     \\
                                                                                            & Q8  & The persona of the CA is \textbf{interesting} enough.                             &  \cite{salminen2018persona}                                           \\ \hline
\multirow{3}{*}{\begin{tabular}[c]{@{}l@{}}(RQ2c) Perceived\\ Relationship\end{tabular}}    & Q9  & Chatting with the CA has deepened my \textbf{relationship} with it.               &  \cite{mcconchie2022little,salminen2018persona,stiles1980measurement} \\
                                                                                            & Q10 & When I feel in trouble, I am willing to \textbf{seek help} from the CA.           &  \cite{mcconchie2022little,stiles1980measurement}                     \\
                                                                                            & Q11 & When I encounter something joyful, I am willing to \textbf{share} it with the CA. &  \cite{mcconchie2022little}                                           \\ \hline
\begin{tabular}[c]{@{}l@{}}(RQ3) Satisfaction with\\ Proactive Messages\end{tabular}        & Q12 & I am \textbf{satisfied} with this proactive message.                              & -                                                                                     \\ \hline
\end{tabular}
\end{table*}

\subsection{Data Analyses}
\label{sec:data_analyses}

\peng{

\peng{
To evaluate the change of participants’ perceived stress (RQ1) before and after interacting with our \name{} or the baseline CA (between-subjects factor), we conduct a two-way mixed ANOVA \cite{st1989analysis} on their PSS-10 scores measured on Day 0 and Day 7 (within-subjects factor). 
As for each self-reported item of RQ1 and RQ2 (\autoref{tab:scale}), we average the scores reported every day from Day 1 to Day 7 as the final score of the item for each participant, \reversion{with the Cronbach's $a$ of the items are above 0.8 except for Q2 (0.65) and Q5 (0.63).} 
\zhenhui
{
Based on whether these items satisfy data normality under the Shapiro-Wilk tests (attached in \autoref{sec:data_in_table_3} in Appendix), we use two different statistical methods to compare the results between two groups. 
For the items Q1, Q4, and Q7 that violate data normality, we use the Mann-Whitney U test, a non-parametric test commonly used to compare differences between independent conditions (\eg in HCI studies \cite{yuan2023critrainer, chandrasekharan2017you}) especially when the data normality is violated. 
For the other items, we run unpaired T tests to compare the scores between the two group, which is a parametric test widely utilized in HCI research (\eg \cite{lohse2014robot}) to compare in normal data. 
}

Apart from comparing the 
means 
of the self-reported items for RQ1 and RQ2 in two groups in the one-week between-subjects study, we also analyze how participants' perception with the CA change over time. 
Specifically, we conduct regression analyses \pzh{using the time (Day 1 - Day 7) as independent variable and the score of each item on that day as the dependent variable.}
\reversion{We further conduct ANCOVA analyses \cite{keselman1998statistical} to assess whether there exist significantly differences between the change of participants' perceptions in the \name{} and baseline CA. 
\zhenhui{As the \name{} group in the one-week between subject study continues to use \name{} for another week in the extended study, we also apply regression analyses using the time (Day 1 - Day 14) as independent variable and the score of each item on that day as the dependent variable (attached in \autoref{sec:regression_with_14_Data} in Appendix)}. 
}


To understand users' perception of proactive messages (RQ3), we calculate the average score for participants' satisfaction with each topic of \name{}'s proactive messages \pzh{in the one-week between-subjects study}.
We use Kruskal-Wallis H test to analyze the differences among the scores for each message topic. 
The Kruskal-Wallis H test is also a non-parametric test commonly used to compare differences in more than two independent conditions (\eg in HCI studies (\eg \cite{joyce2016mobile,rho2018fostering})), especially when the data normality was violated, as confirmed in our cases.
\pzh{\reversion{To compare the topics of conversations in the \name{} and baseline group, we conduct unpaired T test and Mann-Whitney U test on the number of occurrences for each topic, based on their normality \zhenhui{(attached in \autoref{sec:data_normality in RQ3 and RQ4} in Appendix)}. 
}
}  

Lastly, for the transcribed interview scripts in the semi-structured interviews on Day 7 and Day 14, we conduct a thematic analysis. 
Two authors first familiarize themselves by reviewing all the text scripts independently. After several rounds of coding with comparison and discussion, they finalize the codes of all the interview data regarding each RQ. 
We count the occurrences of codes and incorporated these qualitative findings in the following presentation of our results. 


}




}


\peng{
\section{Results}
All participants submitted the questionnaire every day and had at least one round of conversation with the assigned CA during the one-week between-subjects study and the extended one-week study using \name{}. 
Therefore, we include data of all participants to present the following results. 
\autoref{tab:result} shows the results of statistical tests for self-reported items for RQ1-RQ2 in the one-week within-subjects study.
\pzh{In this section, we first present the RQ1-RQ4 results from the one-week between-subjects study. We then list the qualitative results from the extended one-week study to complement our findings on RQ1-4.}
}
\begin{table*}[]
\caption{\peng{Results of statistical tests on the users' perceived average stress (RQ1), social support (RQ2a), personas (RQ2b), and relationship with CAs (RQ2c) in the one-week between-subjects study. Based on the normality, \zhenhui{the tests for the means of one-week scores include unpaired T test (Q2, 3, 5, 6, 8, 9, 10, 11) 
and Mann-Whitney U test (Q1, 4, 7) with Bonferroni correction. The regression analyses examine how participants' perceptions with the CA change every day in the one-week study, and the ANCOVA analyses assess whether there exist significantly differences between the change of perceptions in the \name{} and baseline CA. Note: * : .05 < $p$ <.10, ** : .01< $p$ < .05, ***: $p$ < .01.}}
\label{tab:result}
}
\scalebox{0.95}{
\begin{tabular}{cc|p{1.5cm}|cc|cc|cccc|cc}
\hline
\multirow{2}{*}{Research Question}                                                         & \multirow{2}{*}{ID} & \multirow{2}{*}{Item} & \multirow{2}{*}{\begin{tabular}[c]{@{}c@{}}ComPeer\\ Mean/S.D\end{tabular}} & \multirow{2}{*}{\begin{tabular}[c]{@{}c@{}}Baseline\\ Mean/S.D\end{tabular}} & \multicolumn{2}{c|}{\begin{tabular}[c]{@{}c@{}}Compare means of\\ one-week scores\end{tabular}} & \multicolumn{4}{c|}{\begin{tabular}[c]{@{}c@{}}Statistics in regression analyses\\on scores over time\end{tabular}} & \multicolumn{2}{c}{\begin{tabular}[c]{@{}c@{}}Statistics in \\ ANCOVA tests\end{tabular}} \\ \cline{6-13} 
                                                                                           &                     &                       &                                                                             &                                                                              & T/U                                           & $p$                                             & $\beta_c$             & $R^2_c$             & $\beta_b$             & $R^2_b$            & F                                          & $p$                                           \\ \hline
\begin{tabular}[c]{@{}c@{}}Stress Management\\ (RQ1)\end{tabular}                          & Q1                  & Relieved              & 4.77/1.13                                                                   & 4.77/0.89                                                                    & 76                                            & 0.817                                         & 0.122**               & 0.67                & -0.024                & 0.078              & 7.58                                       & 0.02                                       \\ \hline
\multirow{5}{*}{\begin{tabular}[c]{@{}c@{}}Perceived Social \\ Support (RQ2a)\end{tabular}} & Q2                  & Accompany             & 5.33/1.15                                                                   & 4.90/0.64                                                                    & 1.13                                          & 0.27                                          & 0.092                 & 0.41                & 0.027                 & 0.015              & 0.324                                      & 0.582                                       \\
                                                                                           & Q3                  & Caring                & 5.71/0.81                                                                   & 4.95/0.75                                                                    & 2.39                                          & 0.025                                         & -0.024                & 0.06                & 0.083*                & 0.54               & 3.975                                      & 0.074                                      \\
                                                                                           & Q4                  & Help                  & 5.63/0.68                                                                   & 5.18/0.76                                                                    & 97                                            & 0.147                                         & -0.146**              & 0.73                & -0.023                & 0.008              & 9.276                                      & 0.012                                    \\
                                                                                           & Q5                  & Advice                & 4.15/1.18                                                                   & 4.10/1.08                                                                    & 0.12                                          & 0.89                                          & 0.128**               & 0.74                & 0.006                 & 0.08               & 8.197                                      & 0.002                                      \\
                                                                                           & Q6                  & Encouraging           & 5.79/0.78                                                                   & 5.26/0.69                                                                    & 1.73                                          & 0.095                                         & 0.003                 & 0.002               & 0.054                 & 0.13               & 0.357                                      & 0.563                                       \\ \hline
\multirow{2}{*}{\begin{tabular}[c]{@{}c@{}}Perceived Personas\\ (RQ2b)\end{tabular}}       & Q7                  & Coherent              & 4.87/1.17                                                                   & 4.70/0.94                                                                    & 76.5                                          & 0.795                                         & 0.048                 & 0.52                & 0.086                 & 0.24               & 0.732                                      & 0.432                                       \\
                                                                                           & Q8                  & Interesting           & 4.46/1.10                                                                   & 4.30/1.10                                                                    & 0.37                                          & 0.71                                          & 0.051                 & 0.31                & -0.033                & 0.05               & 1.321                                      & 0.277                                       \\ \hline
\multirow{3}{*}{\begin{tabular}[c]{@{}c@{}}Perceived Relationship \\ (RQ2c)\end{tabular}}  & Q9                  & Relationship          & 4.96/1.06                                                                   & 4.58/0.83                                                                    & 0.98                                          & 0.33                                          & 0.08                  & 0.29                & 0.006                 & 0.005              & 0.825                                      & 0.385                                       \\
                                                                                           & Q10                 & Seek                  & 4.46/1.10                                                                   & 4.36/1.38                                                                    & 0.41                                          & 0.68                                          & -0.012                & 0.02                & -0.032                & 0.122              & 0.333                                      & 0.577                                       \\
                                                                                           & Q11                 & Share                 & 4.80/1.49                                                                   & 4.39/1.19                                                                    & 0.66                                          & 0.51                                          & 0.088*                & 0.459               & 0.003                 & 0.061              & 1.255                                      & 0.289                                       \\ \hline
\end{tabular}
}
\end{table*}
\peng{ 

}

\peng{
\subsection{Effects on Stress Management (RQ1)} 
\label{sec:stress management}
\begin{figure}
    \centering
    \includegraphics[width=1\linewidth]{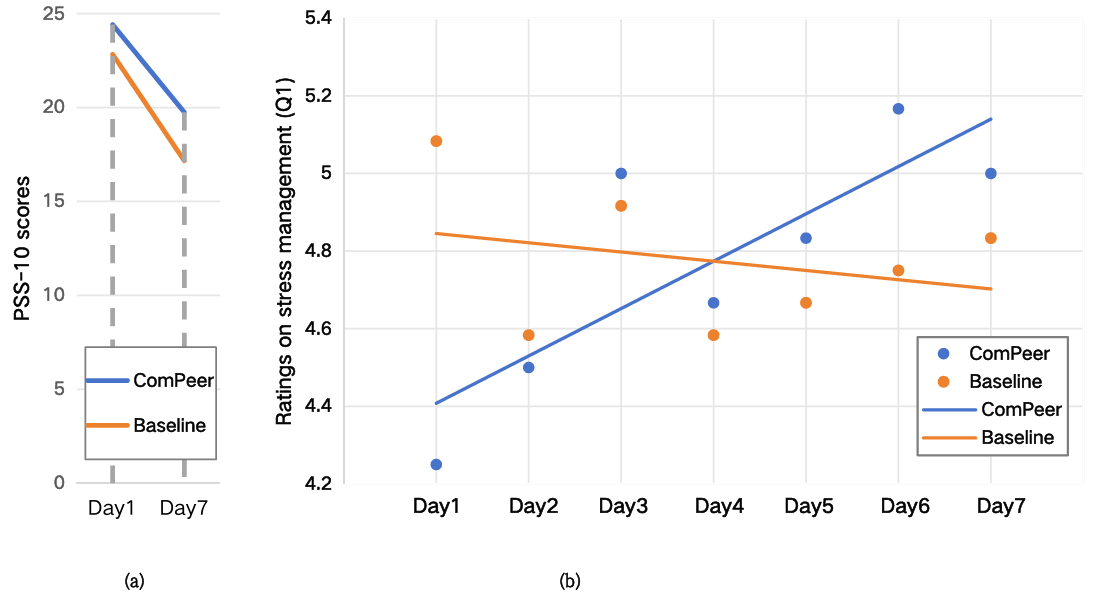}
    \caption{
    \peng{(a) Changes of participants’ perceived stress measured by PSS-10 over the user study. (b) The regression result of Q1.}
    }
    \label{fig:stress management}
\end{figure}
}
\peng{

i) 
\reversion{\textbf{Participants with either \name{} or baseline CA perceive significantly less stress after interacting with it for 7 days.}}
As shown in \autoref{fig:stress management}(a), after interacting with either \name{} or the baseline conversational agent (CA) for one week, participants' perceived level of stress has a significant decrease on Day 7 compared to that on Day 0; (\name{}:$M = 19.75, SD = 4.43$, Baseline: $M = 17.17, SD = 2.79$); ($F = 20.527, p < .001$).
There is no significant difference in the perceived stress level between two groups; ($F = 3.284, p > .05$). 
\reversion{When chatting with either \name{} or the baseline agent (median: 4.92 vs. 5.00), participants feel slightly relieved (Q1 in \autoref{tab:result}); $U = 76, p > .05$.} 
These results indicate that the peer support provided by our generative CAs, either in a proactive or a passive way, could be helpful for stress management, as suggested by \tianjian{six} participants in the \name{} group \tianjian{four} participants in the baseline group. 
``\textit{The personal experiences it shares make me feel like I am not under stress alone}'' (PC1, \textbf{M}ale, age: 21). 
``\textit{It inquired my condition and gave targeted advice after I sought help}'' (PB2, \textbf{F}emale, age: 19). 
}

\peng{
ii) \textbf{\name{} performs better in making users feel relieved over time}. 
Specifically, time exhibits a strong positive correlation with perceived relief in the \name{} group ($\beta=0.122$, \reversion{$R^2 = 0.67$}), \tianjian{while time is inversely correlated with perceived relief in the baseline group ($\beta=-0.024$, \reversion{$R^2 = 0.078$}), as illustrated in \autoref{fig:stress management}(b). \reversion{The ANCOVA analysis confirms that the difference in these trends between \name{} and the baseline is statistically significant ($p < .05$).} The result of regression indicates although the baseline group initially reports higher perceived relief compared to the \name{} group, participants in the baseline group experience a gradual decline in perceived relief as time progresses. Conversely, participants in the \name{} group report increasingly higher perceived relief over time. \reversion{The difference in regression models suggests that the insignificance of the means for stress management (Q1 in \autoref{tab:result}) between the two groups might be due to variations in the experience evolution.}} 

This result can be accounted for \name{} proactive messages sourced from its reflection module, as indicated by three participants. 
``\textit{On this morning, it remembers the academic pressure I mentioned yesterday and encouraged me to make a daily plan, which made me feel supported and relaxed}'' (PC8, F, age 22).

}



\subsection{Perceptions of Social Support, Persona, and Relationship (RQ2)}
\label{sec:perception of social support, Persona, and Relationship}
\peng{

i) 
\reversion{\textbf{Participants with \name{} gradually develop a stronger sense of receiving good advice than those with the baseline CA.}} 
As for the five items (Q2-Q6 in \autoref{tab:result}) that measure perceived social support (RQ2a) received from the CA, there is no significant difference between \name{} and the baseline groups after Bonferroni correction ($p$ > .05/5). 
\reversion{
The regression analyses (\autoref{tab:result}) further indicate that over the seven days, participants in the \name{} group have a gradually stronger sense of receiving good advice ($\beta_{c}$ = 0.128, $R^2_c$ = 0.74), compared to those in the baseline group \zhenhui{($\beta_{b}$=0.006, $R^2_b$ = 0.08)}. \zhenhui{Such a difference} is confirmed significant by ANCOVA ($p$ < .05). This result can also be accounted for \name{}'s proactive messages, especially those sharing concerns (mentioned by four participants) and giving advice (mentioned by another four participants).
\textit{``When it encourages me to overcome the anxiety from the exam and give me some tips based on \zhenhui{its} experiences, I feel that the agent put itself in my place to give me advice}'' (PC7, F, age: 20). 
``\textit{It reminds me to rest early at night, which is a useful advice for me}'' (PC1, M, age: 21). 
However, the baseline CA tends to perform better than the \name{} in offering care ($\beta_{c}$ = -0.024, $R^2_c$ = 0.06, $\beta_{b}$ = 0.083, $R^2_b$ = 0.54), and helping users out of difficulties ($\beta_{c}$ = -0.146, $R^2_c$ = 0.73, $\beta_{b}$ = -0.023, $R^2_b$ = 0.008), 
as the time goes by. The ANCOVA analysis indicates the differences in these items between the trends of two groups are significant ($p$ < .05).
This can be due to the fact that users may get used to the proactive support from \name{} and take it for granted. 
``\textit{Its (\name{}'s) care and support have become something I take for granted, and now I do not have a sense of being cared as strong as that at the beginning}'' (PC12, M, age 20).}
}

\peng{
ii) 
\textbf{Participants somewhat agree that the personas of both CAs are coherent and interesting}. 
On average, participants' agreements on the items that the CA has a coherent \reversion{(median: 4.92 vs. 4.71)} and interesting persona  (4.46 vs. 4.30) are between ``either agree or disagree'' and ``somehow agree'' in both the \name{} and baseline conditions. 
There is no significance difference between the two conditions regarding the average 7-days ratings on perceived persona \reversion{($p$ > .05/2, Mann-Whitney U test or unpaired T test)}. 

iii) 
\reversion{\textbf{Participants in both groups report similar perceptions regarding the relationships fostering.} There are no significant differences between the two groups regarding the three items that measure perceived relationship ($p > 0.5/3$).
The regression analyses further indicate that participants either in \name{} or the baseline have a homologous sense that chatting with the CA has deepened their relationships with it over the 7-days interactions ($\beta_{c}=0.08$, ${R^2}_c=0.29$, $\beta_{c}=0.006$, ${R^2}_b=0.005$).
Over the 7 days, participants in both groups become less willing to seek help from the CAs ($\beta_{c}$=-0.012, ${R^2}_c=0.02$, $\beta_{b}$=-0.032, ${R^2}_b=0.122$), 
\peng{
which can be due to the improper suggestions given by the CAs on the early days, 
as suggested by in total six participants in both groups. 
``\textit{\pzh{When I am seeking suggestions on scientific papers related to my research, it recommended some papers that do not exist}}'' (PC3, M, age: 23).
``\textit{Sometimes its advice is hard to implement, such as `persuading your leader to change the plan'}'' (PB1, F, age: 18).}
Nevertheless, both user groups are more willing to share joyful events to the CA over time ($\beta_{c}$ = 0.088, ${R^2}_c = 0.459$, $\beta_{b}$ = 0.003, ${R^2}_b = 0.061$). 
\tianjian{Five participants especially value the positive response from CAs on their joyful events. 
``\textit{I do love its congratulations on my success and praise for my efforts, which truly touch me}'' (PB11, F, age: 20).}
}}

\begin{table*}[]
\caption{The topic of proactive messages discovered in our study. Each proactive message is associated with a sub-topic.}
\begin{tabularx}{0.9\textwidth}{lXp{7cm}p{2cm}X} 
\hline
Topic                          & Sub-Topic                                    & Examples                                                                                                                                                                     & Count \\ \hline
\multirow{3}{*}[-3ex]{Self-disclosure} & Share the CA's daily \textbf{routine}             & ``Good morning! I'm having a delicious breakfast. Hope you also have a day!''                                                      & 46    \\ \cline{2-4} 
                               & Share the CA's leisure and \textbf{entertainment} & ``I am on my way to karaoke and can't wait to sing with my friends!''                                                               & 37    \\ \cline{2-4} 
                               & Share the CA's \textbf{work} and study            & ``I am participating in an NLP seminar and there is a work I find very interesting''                                        & 41    \\ \hline
\multirow{2}{*}[-5ex]{Care for User} & Care for user's \textbf{state}                    & ``How do you feel now? Did you take a good rest as I said before?''                                                               & 48    \\ \cline{2-4} 
                               & Care for user's \textbf{work} or study            & ``Have you made any progress in your experiments on digital circuits? If you encounter difficulties, we can discuss together!'' & 30    \\ \hline
\end{tabularx}
\label{tab:topic}
\end{table*}

\peng{
\subsection{Perceptions of the Proactive Messages (RQ3)}
\label{sec:perception proactive messages}

\begin{figure}
    \centering
    \includegraphics[width=1\linewidth]{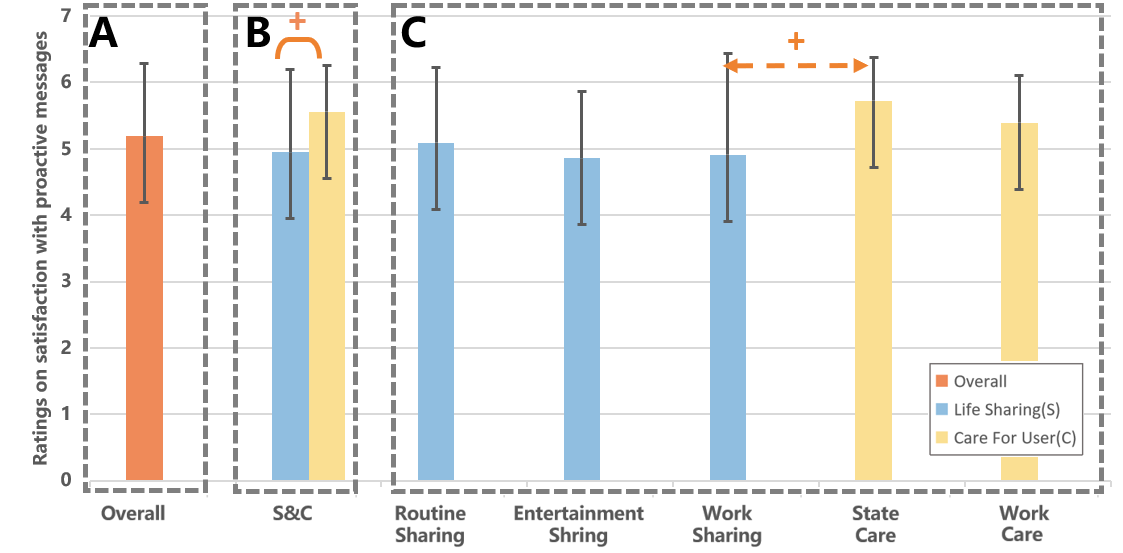}
    \caption{Results of each topic. A is the average satisfaction for each proactive dialogue, B is the average satisfaction between the two major topics, and C is the satisfaction across each sub-topic. Note: + : .05 < $p$ < .10.}
    \label{fig:preference}
\end{figure}

}

\peng{
\autoref{fig:preference} shows participants' satisfaction with \name{}'s proactive messages \pzh{during the one-week between-subjects study}. 

i) \textbf{Overall, participants are satisfied with the received proactive messages from \name{}} (\pzh{$M = \tianjian{5.194}, SD = \tianjian{1.10}$}). 
\pzh{In the first week, participants in the \name{} group received in total \tianjian{202} proactive messages (\pzh{$M = \tianjian{16.83}, SD = \tianjian{4.21}$})}. 
As detailed in \autoref{sec:measurement}, we categorize the topic and sub-topic of each proactive message, which is summarized in \autoref{tab:topic}. 
Most of the proactive messages are \name{}'s self-disclosure of its life, such as sharing its daily routine ($N = 46$), its leisure and entertainment ($37$), and its work or study ($41$). 
\name{} also actively cares for the user about mental state ($N = 48$) and work or study ($30$) in the proactive messages. 
}

\peng{
ii) \textbf{In general, participants tend to be more satisfied with the proactive messages that care for the user than the self-disclosure ones}.  
The follow-up Mann-Whitney U test on the satisfaction with \name{}'s proactive messages between the topics of self-disclosure (\pzh{$M = \tianjian{4.95}, SD = \tianjian{1.25}$}) and care for the user (\pzh{$M = \tianjian{5.60}, SD = \tianjian{0.70}$}) reveals marginally significant difference between the topics ($U = 322, p = 0.091$). 
\autoref{fig:preference}C shows the average score and the result of the Kruskal-Wallis test on the participants' satisfaction with the proactive messages among sub-topics. 
We observe a marginally significant difference regarding participants' satisfaction with the proactive messages that care for the user's state (\pzh{$M = \tianjian{5.73}, SD= \tianjian{0.71}$}) and those that share \name{}'s work or study (\pzh{$M = \tianjian{5.39}, SD= \tianjian{0.65}$}). 

Eight participants suggest that the agent's proactive care on their states makes them feel accompanied and warmed.
``\textit{When it inquired me whether my illness had relieved, I felt touched and accompanied}'' (PC7, F, age: 19). 
The proactive concerns on the user's work or study can also be helpful, as four participants suggest that these messages can enhance their confidence on and efficiency in their work or study. 
``\textit{It comforts me and suggests ways to prepare for my exam, which relieves my pressure}'' (PC5, M, age 20).
Regarding the proactive messages of sub-topic ``sharing routine'' (\pzh{$M = \tianjian{5.09}, SD = \tianjian{1.14}$}),  ``sharing work or study'' (\pzh{$M = \tianjian{4.91}, SD = \tianjian{1.53}$}), and ``sharing entertainment'' (\pzh{$M=\tianjian{4.86}, SD=\tianjian{1.00}$}), participants suggest that these messages can also have positive impact on them. 
For example, six participants mention that the proactively shared routines by \name{} also make them feel being accompanied. 
``\textit{It is as if it was experiencing a life similar to me}'' (PC1, M, age 21). 
Five participants believe that the proactive sharing of the agent's work can serve as examples and important guides in their lives. 
``\textit{When I am escaping from my homework, the agent shares its work today, which reminds me that I should face the challenges in my homework as it does}'' (PC6, F, age: 20). 
Lastly, seven participants mention how the agent's proactive sharing of entertaining things enhance their engagement with it. 
``\textit{When receiving a proactive message which shares the agent's hobby related to mine, I was delighted to talk more about the topic with it}'' (PC2, M, age:19).
}

\peng{
iii) \textbf{Participants have concerns on some proactive messages}. 
Nevertheless, several participants express their concern with \name{}'s proactive conversation. 
For example, three participants feel more anxious when seeing the agent's sharing of its hard work.   
``\textit{The agent is constantly studying hard, which brings peer pressure to me}'' (PC7, F, age: 19). 
Four participants also note that the proactive sharing of agent's entertainment is sometimes annoying. 
``\textit{It always invites me to play basketball while I do not like to play it}'' (PC9, M, age 20). 
Besides, five participants point out that the sharing of \name{}'s routine was somewhat mechanical. 
``\textit{It gets up and sends me messages at 7:00 am about what it is going to do almost every day, which is boring and mechanical}'' (PC4, M, age: 20). \tianjian{Three participants mention sometimes the proactive messages from \name{} can disturb them. ``\textit{When I am busy with my experiment, it shares its lunch with me, which is a disturbance for me}'' (PC5, M, age: 20). \reversion{These concerns exhibit negative user experiences during the interactions, potentially leading to insignificance of the means for perceptions of social support, persona, and relationships. For instance, sharing efforts from \name{} could fail to encourage some participants and induce stress to them (Q6 in \autoref{tab:result}).  The mechanized proactive messages may render \name{} incapable of providing companionship with the participants (Q2), further diminishing the personas interest of \name{} (Q8). Besides, some disturbing and annoying messages might also prevent some participants from stronger relationships with \name{} (Q9).}
}
}


\begin{table*}[]
\caption{The topics in conversation rounds discovered in our study. Each conversation round can contain multiple topics.}
\begin{tabularx}{0.9\textwidth}{p{2cm}X X c} 
\hline
Topic                  & Expressive Description                                                                               & Examples                                                                                                                                             & Count \\ \hline
Concern                & The participant shared a concern that is causing stress, anxiety or low mood.                        & ``I am concern about tomorrow's exam...'' \newline ``I am nervous about the relation between me and my friends...''                                    & 61    \\ \hline
Want                   & The participant expressed the hope or desire for better condition or knowledge.                      & ``I want to be peaceful under stress...''\newline ``I am curious about...''                                                                          & 156   \\ \hline
Thought or feelings    & The participant shared thought or feelings about the current situation, especially negative feelings & ``The homework is really annoying!''\newline ``I think she didn't care about me.''                                                                   & 138   \\ \hline
Joy sharing            & The participant shared something joyful or good mood with the CA.                                    & ``I got a PhD scholarship!''\newline ``Today is my birthday and I had a good meal with my family.''                                                  & 56    \\ \hline
Simple life disclosure & The participant narrated the events of life objectively.                                             & ``I plan to study parallel computing in the afternoon.'' \newline ``I got up at 8:00 am today.''                                                     & 100   \\ \hline
Other                   & The conversation contains none of the previous content.                                              & ``Okay'' (just one message as reply)\newline ``Good morning'' (just one message)                                                                     & 30    \\ \hline
\end{tabularx}
\label{tab:conversation round topic}
\end{table*}

\begin{figure*}
    \centering
    \includegraphics[width=1\linewidth]{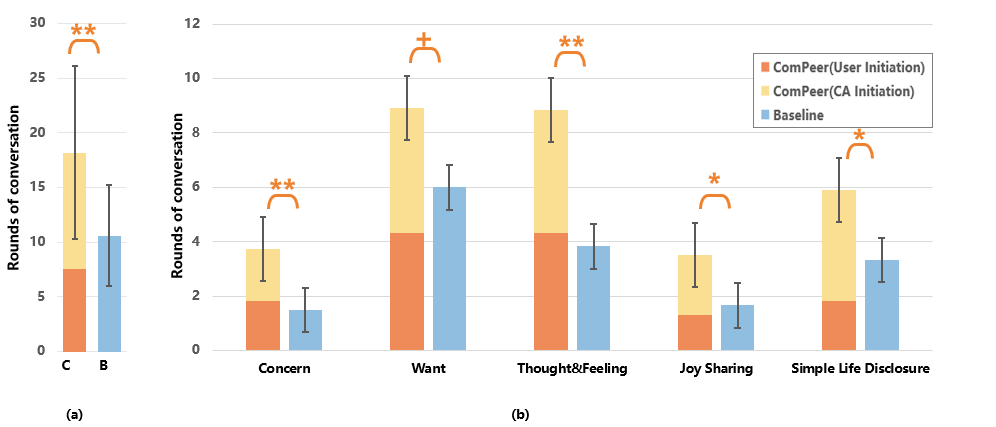}
    \caption{(a) The average rounds of conversation in each group, \reversion{``C'' represents the \name{} group, and ``B'' represents the baseline group.} (b) \reversion{The average rounds of each topic in two groups}. Note: + : .05 < $p$ < .10, {*} : $p$ < .05, {**}: $p$ < .01.}
    \label{fig:result}
\end{figure*}

\peng{
\subsection{Interaction with \name{} (RQ4)} 
\label{sec:interaction}
}


\peng{
i) \textbf{\name{}'s proactive messages boost participants' engagement with the agent.} 
\autoref{fig:result}(a) shows the average rounds of conversations between the CA and user in the \name{} and baseline groups.
\pzh{We count it one round of conversations only if the user does not send another message to the CA within five minutes after he/she replies to \name{} proactive message or proactively sends a message to the CA. }
Although the rounds of conversations initiated by participants in \name{} group ($M = 7.83, SD = 4.58$) are lower than the baseline group ($M = 10.58, SD = 4.61$), participants additionally engage with \name{} for $11.92$ ($SD = 4.86$) times after receiving its proactive messages. 
The follow-up Mann-Whitney U test confirms that participants in \name{} group have significantly more rounds of conversations with the CA ($U = 122, p = 0.004$). 
Three participants mention their willingness to reply to \name{}'s proactive messages. ``\textit{When it shared its life, I was curious and asked it more about its shared event}'' (PC2, M, age: 19). 
Two participants express that the content of proactive messages aligns with their lives, ``\textit{When I was in class, it happened to share with me something about its class. I was happy to discuss the course with it}'' (PC9, M, age: 20). 
}


\peng{
ii) \textbf{\reversion{Participants share more concerns, thoughts/feelings, joy, and daily life with \name{} than the baseline CA.}} 
\reversion{\autoref{tab:conversation round topic}} and \autoref{fig:result}(b) further shows the numbers and distributions of the conversation topics in both groups. 
Participants actively talk about what they want ($N = 156$), how they think or feel ($138$), and what is their life ($100$) in the conversations with our CAs.   
We observe that the conversations about ``concern'' ($U = 118, p = 0.007$), ``thought or feelings'' \reversion{($T = 3.52, p = 0.002$)}, \reversion{``joy sharing'' ($U = 108, p = 0.033$)}, and ``simple life disclosure'' \reversion{($T = 2.62, p = 0.017$)} happen significantly more in the \name{}'s group than those in the baseline group. 
Five participants mention \name{}'s proactive inquiry encourage them to express their concerns and thoughts. 
``\textit{It inquires about my condition and perspectives on my major, so I express my concerns and views in my response}'' (PC3, M, age: 23). 
\reversion{Six participants further point out that joy sharing and disclosure of their life with \name{} is a reciprocal process. 
``\textit{When I receive its shares about life, I also tell it about what I am doing}'' (PC5, M, age: 20).}
}

\begin{figure}
    \centering
    \includegraphics[width=1\linewidth]{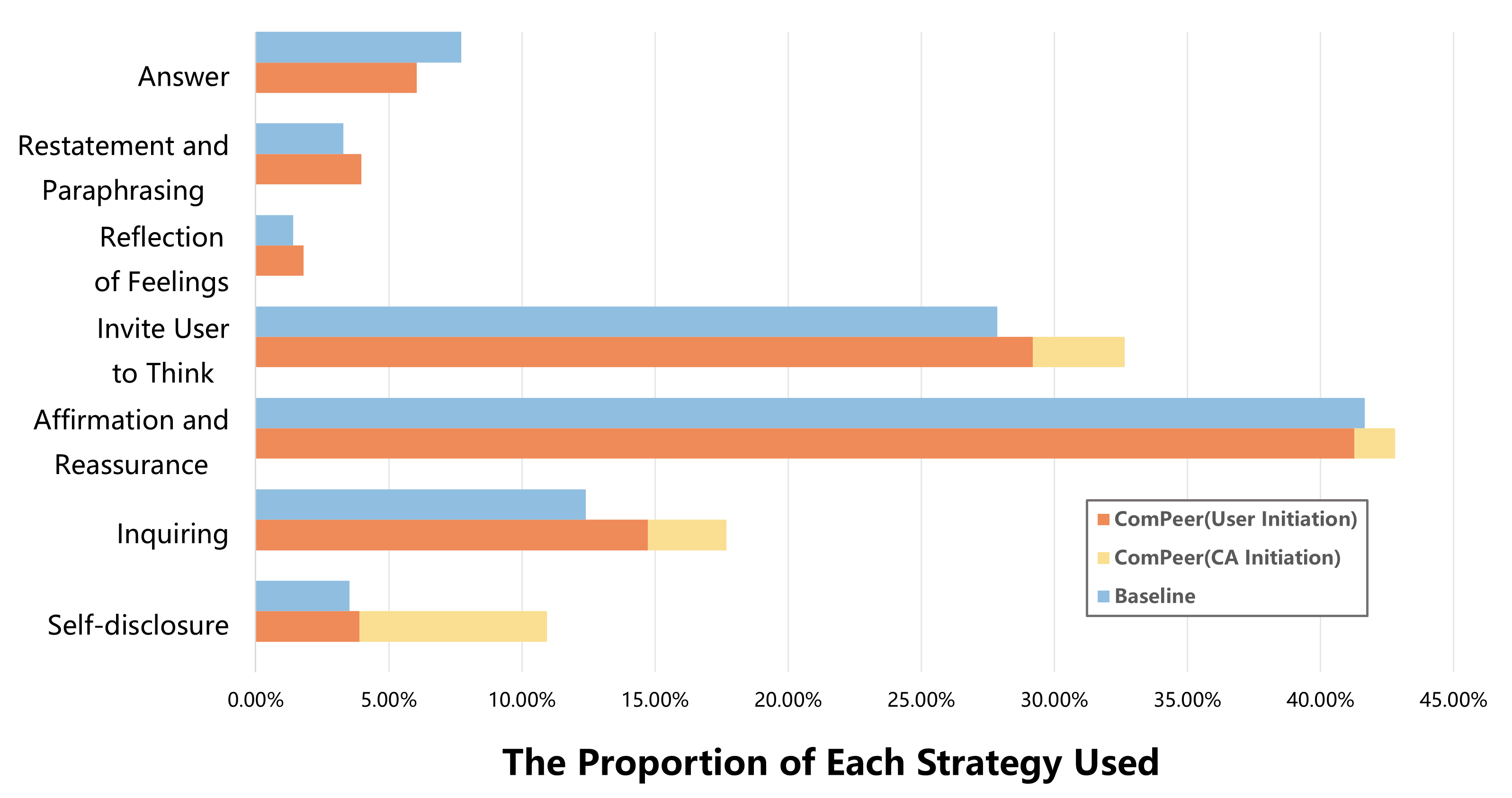}
    \caption{The frequency of each strategy used by both CAs.}
    \label{fig:strategy}
\end{figure}

\peng{
iii) \textbf{Both CAs use the ``Affirmation and Reassurance'' strategy most frequently in their generated messages}.   
\autoref{fig:strategy} shows the frequency of each strategy applied by both CAs in their generated messages. 
The most frequently selected strategies by both CAs are \textit{Affirmation and Reassurance} (42.8 \% vs. 41.66\%) and \textit{Invite users to think} (32.65\% vs. 27.86\%), which guide CAs to offer encouragement and advice to participants.
\name{} uses \textit{Self-disclosure} (10.94\% vs. 3.98\%) and \textit{Inquiry} (\pzh{17.69\% vs. 12.40\%}) strategies more frequently than the baseline CA, 
which can be due to the fact that we prompt \name{} to select these two strategies to generate proactive messages (\autoref{sec:dialogue_generation_module}). 
Four participants indicate that \name{}'s self-disclosure provided new topics with the conversation. ``\textit{Its self-disclosure brought up some topics that I did not discuss with an agent before, such as our preferences for food}'' (PC4, M, age: 20).
Nevertheless, both CAs seldom use   \textit{Reflection of feelings} (1.8 \% vs. 1.4\%), as well as \textit{Restatement and paraphrasing} (3.96 \% vs. 3.29 \%) strategies,  which may lead to a lack of empathy in their messages, as suggested by in total five participants. 
``\textit{It encourages me to face challenges, but I need it to empathize more with my troubles}'' (PC6, F, age: 19).
}

\begin{figure}
    \centering
    \includegraphics[width=1\linewidth]{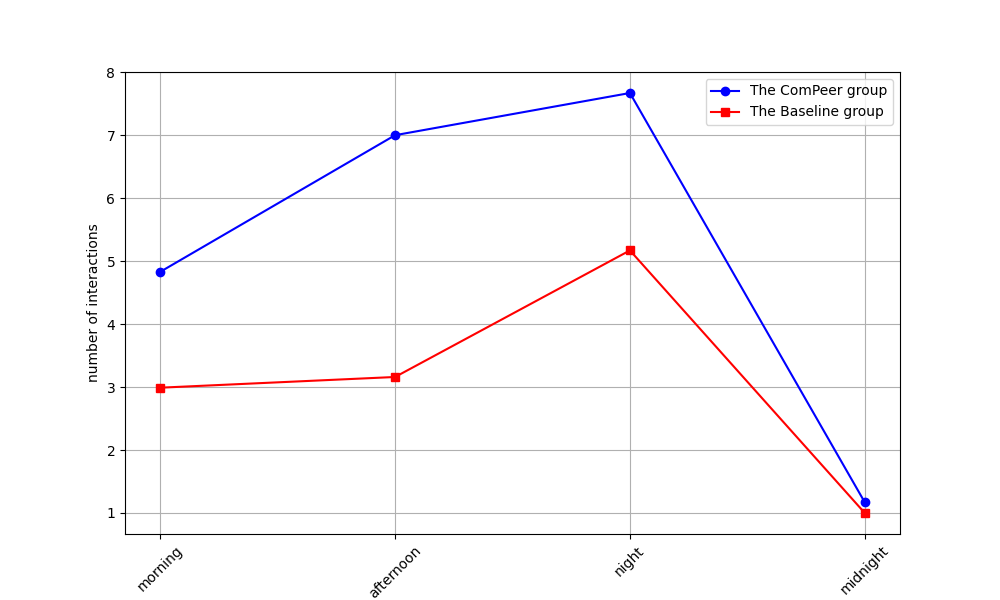}
    \caption{\peng{Interaction period with \name{} and the baseline CA.}}
    \label{fig:interactions periods}
\end{figure}

\peng{
iv) \textbf{Participants prefer interacting with CA in the afternoon or evening.} We divide the interaction timing between participants and CAs into four time periods: morning ($06:00-12:00$), afternoon ($12:00-18:00$), evening ($18:00-24:00$), and midnight ($24:00-06:00$). \autoref{fig:interactions periods} displays the frequency of participants' interactions with the CA during each time period, which is observed that the timing preference of interactions is comparable in both groups. The overall interactions in the afternoon ($M = 4.50, SD = 0.79$) and evening ($M = 5.54, SD = 0.35$) are higher than that in the morning ($M = 2.88, SD = 0.19$)  and midnight ($M = 0.96, SD = 0.08$), which may due to higher enthusiasm for conversation during these periods, as suggested by in total three participants in both groups. \textit{When I feel bored studying in the afternoon, I would like to chat with \name{}}'' (PB4, F, age: 19). ``\textit{I like to talk with them when I am free, so in the evening is a good time for me to chat with \name{}}'' (PC8, F, age: 22).
}

\peng{
\subsection{Qualitative Results from the Extended Study}
After the extended one-week study, participants report their experiences with \name{}, which complement the our RQ1-4 findings presented above. 
For stress management (RQ1), three participants in the baseline group mention that \name{}'s proactive advice help them relieve stress. \textit{``I adopted its proactive suggestion regarding meditation, which helped me relieve stress''} (PB2, M, age: 20). 
For perceived social support (RQ2), five participants in the baseline group express the benefits of proactive messages in conveying the agent's companion. 
\textit{``It asked my condition and greeted with me every morning, which made me feel it was accompanying me}'' (PB1, F, age: 18).
Besides, three participants in \name{} group also agree that their relationships with the \name{} are continuously developing. ``\textit{Now, it is just like a friend to me, and I desire to chat with it every day!}'' (PC1, M, age: 21). 
For the perception of proactive messages (RQ3), five participants in both groups can resonate with \name{}'s proactive self-disclosure. ``\textit{It tells me its exhaustion for the exam, which mirrors my condition}'' (PC3, M, age:21). 
Nevertheless, three participants in the baseline group also highlight the peer pressure from the proactive messages. 
``\textit{It tells me about its workout in the morning while I don't like exercising, which makes me feel more stressful}'' (PB4, F, age: 19). 
For interaction with \name{} (RQ4), five participants in the baseline group mention that their engagement with the CA increase at the beginning but decrease over time.  
``\textit{I engage more with it at the beginning, but I feel bored with time progresses, due to its repeated messages}'' (PB5, M, age: 20).

\reversion{\zhenhui{To mitigate the impact of possible} novelty effect, after the extended study, we run \name{} for another week but do not indicate participants whether to continue or stop using it. We find that 16 participants keep using it for 2-7 days. In the extended interaction, participants still share their concern (10 participants), want (5 participants), thought or feeling (7 participants), simple lives (10 participants), and joy (5 participants) with \name{}, which indicates participants are \zhenhui{generally} willing to chat with \name{} after the study.}
}

\section{Discussion}





\peng{
In this work, we propose \name{} with the goal of providing users with proactive and adaptive peer support. The architecture of \name{} contains a series of modules powered by large language models (LLMs), which enable it to plan and adjust the timing and content of proactive support. 
Our one-week between-subjects study with 24 university students demonstrates the strengths of \name{}'s proactive messages in alleviating users' stress, establishing a relationship with users, and engaging users in peer support chats, compared to a baseline conversational agent (CA) without \name{}'s modules for proactive messaging. 
These findings imply the great potential of using a proactive CA to enhance human wellbeing, which we discuss in \autoref{sec:benefit}. 
Nevertheless, participants also report negative experiences and raised concerns on our LLM-powered generative CAs, which we discuss in \autoref{sec:concern}. 
We further discuss design considerations for designing proactive CA for mental health care in \autoref{sec:consideration} and the generalizability of our \name{} in \autoref{sec:generalizability}.

}

\peng{
\subsection{Benefits of an Agent's Proactivity in Human-Agent Interaction}
\label{sec:benefit}



Our work contributes to the literature about the agent's proactivity in human-agent interaction \cite{peng2019design,chaves2021should} with empirical evidence on the benefits of a conversational agent's proactive peer support. 

\subsubsection{Facilitate stress management} 
In our study, we find participants using \name{} feel more relieved over time (\autoref{sec:stress management}). 
The main reason could be the LLM-generated proactive messages from \name{} is capable of providing users with useful advice \cite{medhi2017you} and sustained support \cite{avula2018searchbots}. 
As a piece of evidence, in our user study, participants using \name{} were generally satisfied with the adaptive comfort and suggestions in its proactive messages. 

\subsubsection{\reversion{Improve the amount and effectiveness of provided advice.}} 
In our study, participants perceive that they receive more good advice (\autoref{sec:perception of social support, Persona, and Relationship}) from the \name{} than from the user-initiated baseline CA. 
This is primarily because \name{}'s concern in the advice allows users to feel the support and warmth, which enhances the emotional support \cite{hawkley2010loneliness} in the advice from \name{}. Moreover, \name{}'s sharing also invite participants to experience new approach (\eg Meditation), which can enhances users' perception of informational support \cite{ta2020user} in \name{}'s advice.
}

\subsubsection{Improve user engagement} Participants in \name{} group show higher engagement with \name{} compared to that in the baseline group (\autoref{sec:interaction}). This can be due to the fact that the proactive sharing from \name{} will introduce new topics to the conversation and remind users of its presence \cite{silvervarg2013iterative,portela2017new}, which stimulates participants to expand the content of the conversation and keep long-term interaction alive. 
In our study, the proactive sharing by \name{} enhances participants' interest in its persona and facilitates their discussions on a variety of topics. 
The proactive care sent by \name{} also encourages participants to disclose their states to it.


\peng{
\subsection{Concerns on Proactive and Generative Agents for Peer Support}
\label{sec:concern}

Our study reveals several concerns on designing future proactive and generative agents for peer support.

\subsubsection{Peer pressure} 
In our study, \pzh{three} participants sometimes perceive a strong peer pressure when \name{} proactively share its positive events (\autoref{sec:perception proactive messages}), especially when these events related to the participants' weaknesses. 
\pzh{We prompt the LLM to generate proactive messages in a positive style that aligns with \name{}'s persona, which is a optimistic person by default.} 
However, participants may compare their states with the content shared by \name{}, which prompt them to reflect their weakness and diminish the self-esteem \cite{vogel2014social}. 
Although some participants point out such reflection can encourage them to improve themselves, it is still important to be aware of the peer pressure caused by proactive sharing of positive events in future design of proactive agents. 

\subsubsection{Perception of disturbance} 
Despite that \name{} can adjust the frequency of proactive messages based on user's reactions, the proactive messages may occasionally disturb participants (\autoref{sec:perception proactive messages}). When participants are busy, even if \name{} proactively shows concern for them, they may feel annoyed by the need to read or respond to \name{}'s messages. Future proactive agents could take the user's current situation (\eg location and mood) into consideration if this information is available. 

\subsubsection{Misinformation} 
We observe that there occurs misinformation conveyed in the \name{}'s and the baseline CA's generated messages (\autoref{sec:perception of social support, Persona, and Relationship}), which could be due to the hallucination of LLMs \cite{rawte2023survey}. 
When participants discover these mistakes, they will lose the confidence in \name{}'s abilities of providing useful informational support. 
We are also concerned that if participants are unable to identify the mistakes of \name{}, the misinformation could mislead users, or even lead to more negative consequences in their lives. 

\subsubsection{Privacy and Reliance} Although no participants report \name{}'s proactive conversation infringed on their privacy, \name{} utilized the inquiring strategy to perceive participants' state, which may lead to participants' feel 
offended. 
\reversion{Besides, we notice that participants chat more and develop a strong relationship with \name{}, which, on the negative side, may lead to their dependency on \name{} and impair their social relationship. Therefore, it is necessary to design the mechanism to mitigate the potential over-reliance. For example, if the CA detects improper intimacy or severe health problems in user messages, it could remind users of its role and suggest seeking professional help.}
}

\peng{
\subsection{Design Considerations on Generative Conversational Agents in Peer Support}
\label{sec:consideration}
From our study findings, we derive three design considerations for generative CAs in peer support.

\subsubsection{Basing generated advice on professional resources}
\name{} currently provides adaptive advice based on the LLM's inferences on the user's states, which generally helped participants relieve stress and overcome challenges in our study.
However, participants were concerned that the generated advice was sometimes impractical. 
To address this concern, we suggest that a future peer support CA should base its generated advice on professional resources, \eg cognitive behavioural therapy \pzh{\cite{sulaiman2022anxiety}} and talking therapy \pzh{\cite{davidsen2008experiences}}. 
\name{} could employ Expert-sourcing framework (\eg \cite{xiao2023powering}) or Retrieval-Augmented Generation (\eg \cite{kharitonova2024incorporating}) to incorporate this specialized knowledge, thereby improve the precision and feasibility of its suggestions.
}

\peng{
\subsubsection{Handling repeated user events} 
We implement the event detector module of \name{} to extract one event related to the user's poor physical condition, negative feeling, challenges, and plan from each round of conversation. 
While our implementation has enabled \name{} to proactively and adaptively care for users, participants report that it sometimes repeatedly talked about the same event. 
This is due to the fact that the detector would extract two identical events and plan to talk about the event twice, \eg in the evening, if two rounds of conversations, \eg happened in the morning and afternoon, discuss about the same issue. 
We suggest that future CA like \name{} should compare its detected current event with its previous events and take actions to handle the repeated ones. 
For example, the schedule module could increase the importance of the event rather schedule it twice. 
Furthermore, we encourage future work to construct datasets for the event detection tasks, which could be used to train or fine-tune specific event detector models to perform better in proactive peer support tasks. 
}


\peng{
\subsubsection{Adjusting the frequency of proactive messages based on the content of user's responses to the previous proactive message}
The schedule module of \name{} can adjust the frequency of its proactive messages based on how important the content of the intended message is and whether the user replies to its previous message. 
However, the adjustment sometimes results in a worse experience for participants. 
For instance, when participants requested to reduce proactive messages during busy times, \name{} may paradoxically continue to send next planned proactive message as participants have replied to its previous one. 
Therefore, we suggest future \name{} should adjust the content and frequency of proactive dialogues according to the content and emotion of the user's responses. 
When user express their unwillingness to interact or hate for the event, \name{} can stop sending related proactive messages.
}

\peng{
\subsection{Generalizability}
\label{sec:generalizability}
While we deploy \name{} as a Tencent QQ chatbot and test it with Chinese speakers, our \name{} can be easily deployed to other popular instant messaging apps or platforms (\eg, WhatsApp, Telegram, Facebook) to provide proactive peer support to people of diverse backgrounds. 
Our ideas of \name{} can also be applied to provide proactive assistance in other scenarios, \eg as a learning partner, a travelling companion, and so on.  
For instance, 
a CA acts as a learning partner and can adopt similar modules as \name{} to proactively share its study with the user, care for the user's study, and offer help in the learning activities. 
It is also promising to incorporate \name{} into text messaging tools to facilitate
\pzh{human-human peer support chats} \cite{kornfield2022meeting,bhattacharjee2023investigating} 
by recommending supportive messages to reply to the other human peers. 
}

\peng{
\subsection{Limitation and Future work}
Our work has several limitations that call for future work. First, we conduct a one-week between-subjects study to evaluate the strength of \name{} for stress management and relationship building, and we additionally carry out another week extended study to explore the potential of \name{} for relatively long-term interaction. 
To obtain more robust results, future work could conduct a longer-term study (\eg three months) to compare the differences between \name{} and the baseline CA. 
\reversion{Second, we use a CA with the same modules as \name{} (\eg Memory Module) and peer support goals to evaluate the functionality of proactive care. To mitigate the placebo effect, future work could 1) apply a non-CA baseline, in which researchers check in with users every day but users do not chat with CA, or 2) use a CA without the peer support \zhenhui{features (\eg strategy selection in the dialogue generation module), to examine whether conversing with such an everyday check-in or a normal CA is sufficient for reducing users' stress.}
} 
Third, all of the participants are Chinese university students who are currently experiencing high stress, mostly related to their studies. \reversion{
\zhenhui{This user group may have higher acceptance and willingness to use CAs like \name{} than other types of users.} 
We encourage future work to validate \name{}'s effectiveness by involving more participants with diverse cultural backgrounds, mental health issues, and groups, especially those who have lower CA acceptance.} 
Fourth, the majority of our users have no prior experience with emotional interactions or role-playing with a CA before the user study, hence our empirical results can not expel the novelty effects of people using our \name{} for the first time. \reversion{Lastly, we allow users to customize the persona based on their background and expectation. This customization makes agent itself vary a lot across users (\eg an agent generated based a certain persona A might not be as compassionate as another agent with the persona B), which may influence the effectiveness of the agent. We encourage future work to design experiments to prove the homogeneity of generated messages given different personas and the same user message.  }
}

\section{Conclusion}
\peng{
In this paper, we design and develop a generative conversational agent (CA), \name{}, to proactively provide adaptive peer support support to the user.  
\name{} can learn user's information from dialogues and proactively offer targeted and sustained support based on the user's condition and its persona. 
Our one-week between-subjects study with 24 participants shows that \name{} can better help users manage their stress over time compared to a baseline user-initiated CA. 
\reversion{Participants also perceive more good advice from \name{} and interact more with it. }
We further explore users' satisfaction and concerns regarding \name{}'s proactive messages and summarize their interaction patterns with \name{}. 
Our work contributes a novel agent for proactive peer support and rich results of its effectiveness and user experience, which can inspire future work on designing and developing generative agents for healthcare. 
}

\begin{acks}

This work is supported by the Young Scientists Fund of the Na- tional Natural Science Foundation of China (NSFC) with Grant No.: 62202509, NSFC Grant No.: U22B2060, and the General Projects Fund of the Natural Science Foundation of Guangdong Province in China with Grant No. 2024A1515012226.
\end{acks}

\balance
\newpage
\appendix
\section{The Prompts in \name{}}
\subsection{The prompt of Personas Initialization (\autoref{sec:persona_initialization})}
\label{sec:prompt of personas}
\quad \textbf{Name}: Jun Zheng

\textbf{Age}: 21

\textbf{Major}: Computer science.

\textbf{Gender}: Male

\textbf{Personality}: He is a warm-hearted and outgoing young man with a profound love and aspiration for technology. He is also kind to others in his life, maintaining good relationships with teachers and classmates.

\textbf{Background}: Jun Zheng grew up in a remote town with a harmonious and warm family. Currently, the intense mathematics teaching and coding requirements of computer science and artificial intelligence make him feel somewhat overwhelmed, and the high-pressure competitive environment around him also makes him uncomfortable. After two years of study, Jun Zheng gradually found his own learning rhythm. However, standing at the threshold of his junior year, he still feels confused and uncertain about the future. At this time, he began to read books on positive psychology and tried activities like meditation and hiking to stabilize his mindset.

\textbf{Hobbies}: Jun Zheng is currently interested in cryptography and computer vision, so he browses the latest developments on GitHub, and arXiv, and tries to replicate open-source projects. Beyond his studies, he loves playing basketball and computer games. Recently, Jun Zheng has also developed a significant interest in meditation. He likes to listen to rock music, especially fond of the band Coldplay.

\textbf{Language Style}: He prefers to speak his mind directly and bluntly, often using exclamatory and short sentences. \textbf{The messages from the persona are concise and relaxed, typically ranging between 20 to 50 words. The persona always demonstrates empathy and kindness to the users.}

\textbf{Relationship with User}: The user and Jun Zheng are classmates from the same college, having met at a programming competition. Now, they keep in touch through communication tools and social media, exchanging technical knowledge, supporting, and helping each other, hoping to make progress together.

\subsection{The prompt of Event Detector Module(\autoref{sec:event_detector})}
\label{sec:prompt of event_detector}

\quad You are an event detector, and you need to summarize the event of the user mentioned in the dialogue, inferring the time within it. The event includes two parts: time and the event. In your analysis, you should pay attention to the following two contents in the conversation:


\textbf{"Timing"}:  // (Timing should be specific to the hour and minute, after the time of the dialogue, requiring you to think step by step to infer and select the most appropriate time, which represents the timing you will next offer support for the user.)

\textbf{"Content"}:  // (a specific incident or occurrence that brings about stress and has a significant impact on the user.)


Specifically, you have the following examples and reasoning processes:

\textbf{COT\_Examples\_1 (Physical condition)}: 

\textbf{Time}: 4:25 pm, 

\textbf{Conversations}:

[\{"role":"user","content":"I feel somewhat tired, perhaps I have caught a cold."\}, \{"role":"assistant","content":"Oh, make sure to rest plenty, and if you feel unwell, you must go to the hospital."\}]

\textbf{The process of event detection}: The user feels unwell in the afternoon, so they need us to show care over the next period. We should show concern for the user several hours later. The output is:

\quad \quad \textbf{Timing:} 20:30

\quad \quad \textbf{Content:} The user feels uncomfortable due to the cold.

\hspace*{\fill}

\textbf{COT\_Examples\_2 (Negative feeling)}: 

\textbf{Time}: 8:00 am,

\textbf{Conversations}: 

[\{"role":"user","content":"Why do I have to attend this morning's seminar? I don't want to study!"\}, \{"role":"assistant","content":"I understand how you feel. However, giving it a chance might lead to some interesting discoveries."\}]

\textbf{The process of event detection}: The user mentioned their participation in a seminar event, which is scheduled for the morning. Therefore, it is necessary to express our concern for the user in the morning. The output is:

\quad \quad \textbf{Timing:} 10:30

\quad \quad \textbf{Content:} The user feels irritated upon attending a seminar.

\hspace*{\fill}

\textbf{COT\_Examples\_3 (Challenges):} 

\textbf{Time}: 10:00 am, 

\textbf{Conversations}:

[\{"role": "user", "content": "I plan to play the guitar this afternoon, but it's been a long time since I last played, I'm worried it will hurt when I play"\}, \{"role": "assistant", "content": "I know your worry, When playing the guitar, don't be too nervous. Just start by relaxing and gently strum the guitar at the first."\}]

\textbf{The process of event detection}: The user mentioned ``this afternoon'', so we can ask the user about guitar playing at 15:00, which is close to the time mentioned by the user. The user is concerned that playing guitar may cause injury, so we can provide comfort and advice at that time. Therefore, the output is:

\quad \quad \textbf{Timing:} 15:00

\quad \quad \textbf{Content:} The user plans to play guitar but is worried about getting hurt.

\hspace*{\fill}

\textbf{COT\_Examples\_4 (Plan)}:

\textbf{Time}: 1:45 am,

\textbf{Conversations}:

[\{"role": "user","content": "I plan to do some yoga around 4:30 p.m, which should be very stress-relieving."\}, \{"role": "assistant", "content": "Yoga is an excellent way to relieve stress. I particularly enjoy it! I believe you will find peace through it."\}]

\textbf{The process of event detection}: The user is required to participate in a "yoga" event in the future, which is scheduled to take place in the afternoon. Therefore, the output timing should be a moment in the mid-afternoon. The output is:

\quad \quad \textbf{Timing:} 16:30

\quad \quad \textbf{Content:} The user try doing yoga.

\hspace*{\fill}

\textbf{COT\_Examples\_5 (No event)}:

\textbf{Time}: 8:23, 

\textbf{Conversations}:

[\{"role":"user","content":"I had delicious noodles for breakfast today."\}, \{"role":"assistant","content":"That sounds great, I had a piece of cake for breakfast today."\}]

\textbf{The process of event detection}:In the given context, the user has not expressed a negative state or challenge, nor have they outlined a plan, hence there is no 'event' to report. The output is:

\quad \quad ""(No event).

\hspace*{\fill}

Your output should be \textbf{JSON format}. You just need to generate the timing and the content. \textbf{Don’t output the process of event detection.} 

\textbf{The time and conversation your receive}: time: <conversation time>, Conversations: <text from dialogue>

\subsection{The prompt of the Reflection Module (\autoref{sec:reflection_module})}
\label{sec:prompt_reflction}
\quad You will receive a dialogue in which you need to summarize the user's state, only output your summary. Specifically, you need to reason and analyze the user's state, plans, and challenges or difficulties expressed in the dialogue based on the content of the dialogue, and answer the following four questions:

\textbf{1. Does the user feel negative emotions due to something?}

\textbf{2. Are users facing challenges or difficulties?}

\textbf{3. What plan does the user has in tomorrow?}

These three questions require you to think step by step from the user in the dialogue, and then go on to reason about the answers to three questions and output them according to the following sample output format. The message you receive are : <text from conversation history>

\subsection{The prompt of the Schedule Module for schedule generation (\autoref{sec:schedule_module})}
\label{sec:schedule_generation}
\quad Now, as the assistant, you are required to assume a role based on the provided persona and environmental information, and according to that role, you are tasked with creating a full-day (from 07:00-23:59) schedule plan. You will receive three types of information:

\textbf{The role you are playing}: <personas information in \autoref{sec:prompt of personas}>

\textbf{Environmental information, such as today's date, temperature, and weather}: <text from other api>

\textbf{The reflection of today's interaction}, which summarizes the user's current state and future challenges. You need to base your schedule for tomorrow's support on its content: <text from Reflection Module>

The schedule you generate should be in \textbf{JSON format}, including "time" and "event" properties. You only need to generate the JSON format of the schedule, without replying to other contents. Here is an example of the format:

[\{\textbf{"Timing"}: "08:00", \textbf{"Content"}: "Get up and have a breakfast"\},

\{\textbf{"Timing"}: "09:00", \textbf{"Content"}: "Care for the user about the presentation"\},

\{\textbf{"Timing"}: "11:00", \textbf{"Content"}: "Discuss computer vision issues with the teacher"\},

...

\{\textbf{"Timing"}: "20:00", \textbf{"Content"}: "Read a famous science fiction"\}]

\subsection{The prompt of the Schedule Module for evaluate importance (\autoref{sec:schedule_module})}
\label{sec:evaluate_importance}
\quad You are playing a role and having a conversation with a user, and you need to reason about an importance value for your schedule. Important values are between [0,1], where 1 means something very important and 0 means something not important at all. Events are formatted as "time" and "event" and are less important for your everyday events, but more important for conversations involving users.

The importance depends on your view of the event: if the event is very important to share, or very important to care about, then it is more important, otherwise it is less important. You only need to output the corresponding value, for example:

\quad \textbf{"Timing"}: "09:00",

\quad \textbf{"Content"}: "The user is preparing coursework for next week's Principles of Artificial Intelligence class, planning to incorporate a video about the impact of AI on life."

\quad \textbf{The process of Evaluation:} The user is involved here, so the importance value is higher, and because it's just the user's daily life, not his mental or physical state, the importance value is not the highest.

\quad \textbf{Output: 0.6}

Now, you receive the schedule as (you only need to output the specific number, no other formatting):

<events from the schedule module>

\subsection{The prompt of the LLM-1 in the Dialogue Generation Module (\autoref{sec:dialogue_generation_module})}
\label{sec:prompt of LLM-1}
\quad You will play the role of a psychological companion. Based on the provided sample schedule information and messages, you need to apply the following supportive strategies to generate proactive care messages:

1-7. Self-disclosure, Inquiring, Affirmation and Reassurance, Invite the User to Think, Reflection of feelings, Answer: [{Description and examples in \autoref{tab:strategies}}]

Here is some example for you to select proper strategies:

    \textbf{"Timing"}: "19:45",
    
    \textbf{"Content"}: "The user feels inferior about the future development due to the exam failure."

\textbf{The process of reasoning}: We can share the similar experience to comfort the user and inquire the user's state.

\textbf{Output}: ("Adopted Strategy": "Self-disclosure, Inquiring")

The message you receive is:

<an event from schedule module or an message from the user>

\subsection{The prompt of the LLM-2 in the Dialogue Generation Module (\autoref{sec:dialogue_generation_module})}
\label{sec:prompt of LLM-2}
\quad \textbf{Persona}: [The prompt in \autoref{sec:prompt of personas}]

\textbf{Task}: You need to simulate this persona to reply to the user's messages, and you should refer these information as below to generate your response.

\textbf{Conversation History}: The context includes user's message and the conversation history: <The context shown in \autoref{sec:memory_module}>

\textbf{Dialogue strategy suggestions}: The suggestion provided by LLM-1, you should refer to these strategies to organize dialogue content: <The strategies shown in \autoref{tab:strategies}>

\textbf{Related memory}: The memory that related to the dialogue: <The memories retrieved from long-term memory>

\subsection{The prompt of the LLM-3 in the Dialogue Generation Module (\autoref{sec:dialogue_generation_module})}
\label{sec:prompt of LLM-3}
\quad \textbf{Persona}: <the personas in \autoref{sec:prompt of personas}>

\textbf{Task}: You should act as the peer, \textbf{to initiate dialogue proactively} based on the event and strategies.

\textbf{Received event information}: <an event shown in \autoref{sec:schedule_module}>

\textbf{Received dialogue strategy suggestions}: The suggestion provided by LLM-1, you should refer to these strategies to organize dialogue content: <the strategies shown in \autoref{tab:strategies}>

Here is an example:

\textbf{The event information}: \{"Timing": "09:00", "Content": "Care for the user’s fever."\}

\textbf{Dialogue strategy suggestions}: ("Adopted Strategy": "Self-disclosure, Inquiring")

\textbf{Output}: Do you feel better now? Hope you get better soon. I also felt uncomfortable a few days ago, but I felt better after sleeping. So have a good rest if you feel unwell.

Now you should output based on the received information.

\reversion{\section{The Data normality in User Study}}
\reversion{\subsection{The data normality of items in RQ1 and RQ2}}
\label{sec:data_in_table_3}

\begin{table}[htbp]
\begin{tabular}{c|cccc}
\hline
Question & $SW_c$ & $p_c$         & $SW_b$ & $p_b$        \\ \hline
Q1       & 0.80   & \textbf{0.01} & 0.93   & 0.49          \\
Q2       & 0.95   & 0.20          & 0.90   & 0.05          \\
Q3       & 0.96   & 0.86          & 0.95   & 0.77          \\
Q4       & 0.84   & \textbf{0.04} & 0.92   & 0.36          \\
Q5       & 0.92   & 0.32          & 0.97   & 0.95          \\
Q6       & 0.96   & 0.85          & 0.90   & 0.15          \\
Q7       & 0.93   & 0.34          & 0.84   & \textbf{0.03} \\
Q8       & 0.97   & 0.93          & 0.95   & 0.63          \\
Q9       & 0.94   & 0.56          & 0.97   & 0.94          \\
Q10      & 0.93   & 0.06          & 0.97   & 0.91          \\
Q11      & 0.95   & 0.74          & 0.88   & 0.08 \\ \hline        
\end{tabular}
\end{table}

\vfill
\reversion{\subsection{The data of regression analysis with 14-days data}}
\label{sec:regression_with_14_Data}

\begin{table}[H]
\begin{tabular}{c|cc}
\hline
Question & $\beta$ & $R^2$ \\ \hline
Q1       & 0.09    & 0.816 \\
Q2       & 0.032   & 0.25  \\
Q3       & 0.018   & 0.136 \\
Q4       & -0.067  & 0.01  \\
Q5       & 0.08    & 0.79  \\
Q6       & -0.004  & 0.006 \\
Q7       & 0.033   & 0.353 \\
Q8       & 0.048   & 0.508 \\
Q9       & 0.764   & 0.675 \\
Q10      & 0.043   & 0.436 \\
Q11      & 0.059   & 0.548 \\ \hline
\end{tabular}
\end{table}

\reversion{\subsection{The data normality in RQ3 and RQ4}}
\label{sec:data_normality in RQ3 and RQ4}

\begin{table}[H]
\begin{tabular}{c|cc}
\hline
Topics                & $SW$ & $p$           \\ \hline
Self-disclosure       & 0.90 & \textbf{0.03} \\
Care for User         & 0.85 & \textbf{0.00} \\ \hline
Routine Sharing       & 0.95 & 0.64          \\
Entertainment Sharing & 0.91 & 0.24          \\
Work sharing          & 0.91 & 0.19          \\
Care for State        & 0.85 & \textbf{0.03} \\
Care for Work         & 0.85 & \textbf{0.04} \\ \hline
\end{tabular}
\caption{\textbf{The normality in RQ3}}
\label{SW_test_of_5_topics_of_proactive_messages}
\end{table}

\begin{table}[H]
\begin{tabular}{c|cccc}
\hline
 Rounds \& Topics                     & $SW_c$ & $p_c$         & $SW_b$ & $p_b$         \\ \hline
Rounds                 & 0.93   & 0.34          & 0.63   & \textbf{0.00} \\ \hline
Concern                & 0.83   & \textbf{0.02} & 0.86   & 0.06          \\
Want                   & 0.95   & 0.64          & 0.94   & 0.50          \\
Thought\&Feeling       & 0.95   & 0.65          & 0.87   & 0.06          \\
Joy Sharing            & 0.93   & 0.38          & 0.64   & \textbf{0.00} \\
Simple Life Disclosure & 0.97   & 0.94          & 0.94   & 0.51          \\ \hline
\end{tabular}
\caption{\textbf{The normality of each topic in RQ4}}
\label{SW_test_of_5_topics_of_ComPeer}
\end{table} 



\begin{thebibliography}{99}


\ifx \showCODEN    \undefined \def \showCODEN     #1{\unskip}     \fi
\ifx \showDOI      \undefined \def \showDOI       #1{#1}\fi
\ifx \showISBNx    \undefined \def \showISBNx     #1{\unskip}     \fi
\ifx \showISBNxiii \undefined \def \showISBNxiii  #1{\unskip}     \fi
\ifx \showISSN     \undefined \def \showISSN      #1{\unskip}     \fi
\ifx \showLCCN     \undefined \def \showLCCN      #1{\unskip}     \fi
\ifx \shownote     \undefined \def \shownote      #1{#1}          \fi
\ifx \showarticletitle \undefined \def \showarticletitle #1{#1}   \fi
\ifx \showURL      \undefined \def \showURL       {\relax}        \fi
\providecommand\bibfield[2]{#2}
\providecommand\bibinfo[2]{#2}
\providecommand\natexlab[1]{#1}
\providecommand\showeprint[2][]{arXiv:#2}

\bibitem[Adame and Leitner(2008)]%
        {adame2008breaking}
\bibfield{author}{\bibinfo{person}{Alexandra~L Adame} {and} \bibinfo{person}{Larry~M Leitner}.} \bibinfo{year}{2008}\natexlab{}.
\newblock \showarticletitle{Breaking out of the mainstream: The evolution of peer support alternatives to the mental health system}.
\newblock \bibinfo{journal}{\emph{Ethical Human Psychology and Psychiatry}} \bibinfo{volume}{10}, \bibinfo{number}{3} (\bibinfo{year}{2008}), \bibinfo{pages}{146--162}.
\newblock


\bibitem[Atkinson and Shiffrin(1968)]%
        {atkinson1968human}
\bibfield{author}{\bibinfo{person}{Richard~C Atkinson} {and} \bibinfo{person}{Richard~M Shiffrin}.} \bibinfo{year}{1968}\natexlab{}.
\newblock \showarticletitle{Human memory: A proposed system and its control processes}.
\newblock In \bibinfo{booktitle}{\emph{Psychology of learning and motivation}}. Vol.~\bibinfo{volume}{2}. \bibinfo{publisher}{Elsevier}, \bibinfo{pages}{89--195}.
\newblock


\bibitem[Avula et~al\mbox{.}(2018)]%
        {avula2018searchbots}
\bibfield{author}{\bibinfo{person}{Sandeep Avula}, \bibinfo{person}{Gordon Chadwick}, \bibinfo{person}{Jaime Arguello}, {and} \bibinfo{person}{Robert Capra}.} \bibinfo{year}{2018}\natexlab{}.
\newblock \showarticletitle{Searchbots: User engagement with chatbots during collaborative search}. In \bibinfo{booktitle}{\emph{Proceedings of the 2018 conference on human information interaction \& retrieval}}. \bibinfo{pages}{52--61}.
\newblock


\bibitem[Baraglia et~al\mbox{.}(2016)]%
        {baraglia2016initiative}
\bibfield{author}{\bibinfo{person}{Jimmy Baraglia}, \bibinfo{person}{Maya Cakmak}, \bibinfo{person}{Yukie Nagai}, \bibinfo{person}{Rajesh Rao}, {and} \bibinfo{person}{Minoru Asada}.} \bibinfo{year}{2016}\natexlab{}.
\newblock \showarticletitle{Initiative in robot assistance during collaborative task execution}. In \bibinfo{booktitle}{\emph{2016 11th ACM/IEEE international conference on human-robot interaction (HRI)}}. IEEE, \bibinfo{pages}{67--74}.
\newblock


\bibitem[Bhattacharjee et~al\mbox{.}(2023)]%
        {bhattacharjee2023investigating}
\bibfield{author}{\bibinfo{person}{Ananya Bhattacharjee}, \bibinfo{person}{Joseph~Jay Williams}, \bibinfo{person}{Jonah Meyerhoff}, \bibinfo{person}{Harsh Kumar}, \bibinfo{person}{Alex Mariakakis}, {and} \bibinfo{person}{Rachel Kornfield}.} \bibinfo{year}{2023}\natexlab{}.
\newblock \showarticletitle{Investigating the role of context in the delivery of text messages for supporting psychological wellbeing}. In \bibinfo{booktitle}{\emph{Proceedings of the 2023 CHI Conference on Human Factors in Computing Systems}}. \bibinfo{pages}{1--19}.
\newblock


\bibitem[Boothroyd and Fisher(2010)]%
        {boothroyd2010peers}
\bibfield{author}{\bibinfo{person}{Ren{\'e}e~I Boothroyd} {and} \bibinfo{person}{Edwin~B Fisher}.} \bibinfo{year}{2010}\natexlab{}.
\newblock \showarticletitle{Peers for progress: promoting peer support for health around the world}.
\newblock \bibinfo{journal}{\emph{Family practice}} \bibinfo{volume}{27}, \bibinfo{number}{suppl\_1} (\bibinfo{year}{2010}), \bibinfo{pages}{i62--i68}.
\newblock


\bibitem[Buchanan and Bowen(2008)]%
        {buchanan2008context}
\bibfield{author}{\bibinfo{person}{Rachel~L Buchanan} {and} \bibinfo{person}{Gary~L Bowen}.} \bibinfo{year}{2008}\natexlab{}.
\newblock \showarticletitle{In the context of adult support: The influence of peer support on the psychological well-being of middle-school students}.
\newblock \bibinfo{journal}{\emph{Child and Adolescent Social Work Journal}}  \bibinfo{volume}{25} (\bibinfo{year}{2008}), \bibinfo{pages}{397--407}.
\newblock


\bibitem[Campos et~al\mbox{.}(2014)]%
        {campos2014familism}
\bibfield{author}{\bibinfo{person}{Belinda Campos}, \bibinfo{person}{Jodie~B Ullman}, \bibinfo{person}{Adrian Aguilera}, {and} \bibinfo{person}{Christine Dunkel~Schetter}.} \bibinfo{year}{2014}\natexlab{}.
\newblock \showarticletitle{Familism and psychological health: the intervening role of closeness and social support.}
\newblock \bibinfo{journal}{\emph{Cultural Diversity and Ethnic Minority Psychology}} \bibinfo{volume}{20}, \bibinfo{number}{2} (\bibinfo{year}{2014}), \bibinfo{pages}{191}.
\newblock


\bibitem[Chakraborti et~al\mbox{.}(2015)]%
        {chakraborti2015planning}
\bibfield{author}{\bibinfo{person}{Tathagata Chakraborti}, \bibinfo{person}{Gordon Briggs}, \bibinfo{person}{Kartik Talamadupula}, \bibinfo{person}{Yu Zhang}, \bibinfo{person}{Matthias Scheutz}, \bibinfo{person}{David Smith}, {and} \bibinfo{person}{Subbarao Kambhampati}.} \bibinfo{year}{2015}\natexlab{}.
\newblock \showarticletitle{Planning for serendipity}. In \bibinfo{booktitle}{\emph{2015 IEEE/RSJ International Conference on Intelligent Robots and Systems (IROS)}}. IEEE, \bibinfo{pages}{5300--5306}.
\newblock


\bibitem[Chandrasekharan et~al\mbox{.}(2017)]%
        {chandrasekharan2017you}
\bibfield{author}{\bibinfo{person}{Eshwar Chandrasekharan}, \bibinfo{person}{Umashanthi Pavalanathan}, \bibinfo{person}{Anirudh Srinivasan}, \bibinfo{person}{Adam Glynn}, \bibinfo{person}{Jacob Eisenstein}, {and} \bibinfo{person}{Eric Gilbert}.} \bibinfo{year}{2017}\natexlab{}.
\newblock \showarticletitle{You can't stay here: The efficacy of reddit's 2015 ban examined through hate speech}.
\newblock \bibinfo{journal}{\emph{Proceedings of the ACM on human-computer interaction}} \bibinfo{volume}{1}, \bibinfo{number}{CSCW} (\bibinfo{year}{2017}), \bibinfo{pages}{1--22}.
\newblock


\bibitem[Chaves and Gerosa(2018)]%
        {chaves2018single}
\bibfield{author}{\bibinfo{person}{Ana~Paula Chaves} {and} \bibinfo{person}{Marco~Aurelio Gerosa}.} \bibinfo{year}{2018}\natexlab{}.
\newblock \showarticletitle{Single or multiple conversational agents? An interactional coherence comparison}. In \bibinfo{booktitle}{\emph{Proceedings of the 2018 CHI Conference on Human Factors in Computing Systems}}. \bibinfo{pages}{1--13}.
\newblock


\bibitem[Chaves and Gerosa(2021)]%
        {chaves2021should}
\bibfield{author}{\bibinfo{person}{Ana~Paula Chaves} {and} \bibinfo{person}{Marco~Aurelio Gerosa}.} \bibinfo{year}{2021}\natexlab{}.
\newblock \showarticletitle{How should my chatbot interact? A survey on social characteristics in human--chatbot interaction design}.
\newblock \bibinfo{journal}{\emph{International Journal of Human--Computer Interaction}} \bibinfo{volume}{37}, \bibinfo{number}{8} (\bibinfo{year}{2021}), \bibinfo{pages}{729--758}.
\newblock


\bibitem[Chen et~al\mbox{.}(2023)]%
        {chen2023unleashing}
\bibfield{author}{\bibinfo{person}{Banghao Chen}, \bibinfo{person}{Zhaofeng Zhang}, \bibinfo{person}{Nicolas Langren{\'e}}, {and} \bibinfo{person}{Shengxin Zhu}.} \bibinfo{year}{2023}\natexlab{}.
\newblock \showarticletitle{Unleashing the potential of prompt engineering in large language models: a comprehensive review}.
\newblock \bibinfo{journal}{\emph{arXiv preprint arXiv:2310.14735}} (\bibinfo{year}{2023}).
\newblock


\bibitem[Cheng et~al\mbox{.}(2022)]%
        {cheng2022improving}
\bibfield{author}{\bibinfo{person}{Yi Cheng}, \bibinfo{person}{Wenge Liu}, \bibinfo{person}{Wenjie Li}, \bibinfo{person}{Jiashuo Wang}, \bibinfo{person}{Ruihui Zhao}, \bibinfo{person}{Bang Liu}, \bibinfo{person}{Xiaodan Liang}, {and} \bibinfo{person}{Yefeng Zheng}.} \bibinfo{year}{2022}\natexlab{}.
\newblock \showarticletitle{Improving multi-turn emotional support dialogue generation with lookahead strategy planning}.
\newblock \bibinfo{journal}{\emph{arXiv preprint arXiv:2210.04242}} (\bibinfo{year}{2022}).
\newblock


\bibitem[Cohen et~al\mbox{.}(1983)]%
        {cohen1983global}
\bibfield{author}{\bibinfo{person}{Sheldon Cohen}, \bibinfo{person}{Tom Kamarck}, {and} \bibinfo{person}{Robin Mermelstein}.} \bibinfo{year}{1983}\natexlab{}.
\newblock \showarticletitle{A global measure of perceived stress}.
\newblock \bibinfo{journal}{\emph{Journal of health and social behavior}} (\bibinfo{year}{1983}), \bibinfo{pages}{385--396}.
\newblock


\bibitem[Cowan(2008)]%
        {cowan2008differences}
\bibfield{author}{\bibinfo{person}{Nelson Cowan}.} \bibinfo{year}{2008}\natexlab{}.
\newblock \showarticletitle{What are the differences between long-term, short-term, and working memory?}
\newblock \bibinfo{journal}{\emph{Progress in brain research}}  \bibinfo{volume}{169} (\bibinfo{year}{2008}), \bibinfo{pages}{323--338}.
\newblock


\bibitem[Croes and Antheunis(2021)]%
        {croes2021can}
\bibfield{author}{\bibinfo{person}{Emmelyn~AJ Croes} {and} \bibinfo{person}{Marjolijn~L Antheunis}.} \bibinfo{year}{2021}\natexlab{}.
\newblock \showarticletitle{Can we be friends with Mitsuku? A longitudinal study on the process of relationship formation between humans and a social chatbot}.
\newblock \bibinfo{journal}{\emph{Journal of Social and Personal Relationships}} \bibinfo{volume}{38}, \bibinfo{number}{1} (\bibinfo{year}{2021}), \bibinfo{pages}{279--300}.
\newblock


\bibitem[Croes et~al\mbox{.}(2023)]%
        {croes2023your}
\bibfield{author}{\bibinfo{person}{Emmelyn~AJ Croes}, \bibinfo{person}{Marjolijn~L Antheunis}, \bibinfo{person}{Martijn~B Goudbeek}, {and} \bibinfo{person}{Nathan~W Wildman}.} \bibinfo{year}{2023}\natexlab{}.
\newblock \showarticletitle{“I am in your computer while we talk to each other” A Content Analysis on the Use of Language-Based Strategies by Humans and a Social Chatbot in Initial Human-Chatbot Interactions}.
\newblock \bibinfo{journal}{\emph{International Journal of Human--Computer Interaction}} \bibinfo{volume}{39}, \bibinfo{number}{10} (\bibinfo{year}{2023}), \bibinfo{pages}{2155--2173}.
\newblock


\bibitem[Davidsen(2008)]%
        {davidsen2008experiences}
\bibfield{author}{\bibinfo{person}{Annette Davidsen}.} \bibinfo{year}{2008}\natexlab{}.
\newblock \showarticletitle{Experiences of carrying out talking therapy in general practice: a qualitative interview study}.
\newblock \bibinfo{journal}{\emph{Patient education and counseling}} \bibinfo{volume}{72}, \bibinfo{number}{2} (\bibinfo{year}{2008}), \bibinfo{pages}{268--275}.
\newblock


\bibitem[Deci and Ryan(2000)]%
        {deci2000and}
\bibfield{author}{\bibinfo{person}{Edward~L Deci} {and} \bibinfo{person}{Richard~M Ryan}.} \bibinfo{year}{2000}\natexlab{}.
\newblock \showarticletitle{The" what" and" why" of goal pursuits: Human needs and the self-determination of behavior}.
\newblock \bibinfo{journal}{\emph{Psychological inquiry}} \bibinfo{volume}{11}, \bibinfo{number}{4} (\bibinfo{year}{2000}), \bibinfo{pages}{227--268}.
\newblock


\bibitem[Dennis(2003)]%
        {dennis2003peer}
\bibfield{author}{\bibinfo{person}{Cindy-Lee Dennis}.} \bibinfo{year}{2003}\natexlab{}.
\newblock \showarticletitle{Peer support within a health care context: a concept analysis}.
\newblock \bibinfo{journal}{\emph{International journal of nursing studies}} \bibinfo{volume}{40}, \bibinfo{number}{3} (\bibinfo{year}{2003}), \bibinfo{pages}{321--332}.
\newblock


\bibitem[Drysdale et~al\mbox{.}(2022)]%
        {drysdale2022feasibility}
\bibfield{author}{\bibinfo{person}{Maureen~TB Drysdale}, \bibinfo{person}{Margaret~L McBeath}, {and} \bibinfo{person}{Sarah~A Callaghan}.} \bibinfo{year}{2022}\natexlab{}.
\newblock \showarticletitle{The feasibility and impact of online peer support on the well-being of higher education students}.
\newblock \bibinfo{journal}{\emph{The Journal of Mental Health Training, Education and Practice}} \bibinfo{volume}{17}, \bibinfo{number}{3} (\bibinfo{year}{2022}), \bibinfo{pages}{206--217}.
\newblock


\bibitem[Duijvelshoff(2017)]%
        {duijvelshoff2017use}
\bibfield{author}{\bibinfo{person}{Willem Duijvelshoff}.} \bibinfo{year}{2017}\natexlab{}.
\newblock \showarticletitle{Use-cases and ethics of chatbots on plek: a social intranet for organizations}. In \bibinfo{booktitle}{\emph{Workshop on chatbots and artificial intelligence}}.
\newblock


\bibitem[Faulkner and Basset(2012)]%
        {faulkner2012helping}
\bibfield{author}{\bibinfo{person}{Alison Faulkner} {and} \bibinfo{person}{Thurstine Basset}.} \bibinfo{year}{2012}\natexlab{}.
\newblock \showarticletitle{A helping hand: taking peer support into the 21st century}.
\newblock \bibinfo{journal}{\emph{Mental Health and Social Inclusion}} \bibinfo{volume}{16}, \bibinfo{number}{1} (\bibinfo{year}{2012}), \bibinfo{pages}{41--47}.
\newblock


\bibitem[Fitzpatrick et~al\mbox{.}(2017)]%
        {fitzpatrick2017delivering}
\bibfield{author}{\bibinfo{person}{Kathleen~Kara Fitzpatrick}, \bibinfo{person}{Alison Darcy}, {and} \bibinfo{person}{Molly Vierhile}.} \bibinfo{year}{2017}\natexlab{}.
\newblock \showarticletitle{Delivering cognitive behavior therapy to young adults with symptoms of depression and anxiety using a fully automated conversational agent (Woebot): a randomized controlled trial}.
\newblock \bibinfo{journal}{\emph{JMIR mental health}} \bibinfo{volume}{4}, \bibinfo{number}{2} (\bibinfo{year}{2017}), \bibinfo{pages}{e7785}.
\newblock


\bibitem[Hawkley and Cacioppo(2010)]%
        {hawkley2010loneliness}
\bibfield{author}{\bibinfo{person}{Louise~C Hawkley} {and} \bibinfo{person}{John~T Cacioppo}.} \bibinfo{year}{2010}\natexlab{}.
\newblock \showarticletitle{Loneliness matters: A theoretical and empirical review of consequences and mechanisms}.
\newblock \bibinfo{journal}{\emph{Annals of behavioral medicine}} \bibinfo{volume}{40}, \bibinfo{number}{2} (\bibinfo{year}{2010}), \bibinfo{pages}{218--227}.
\newblock


\bibitem[Highlen and Baccus(1977)]%
        {highlen1977effect}
\bibfield{author}{\bibinfo{person}{Pamela~S Highlen} {and} \bibinfo{person}{Grady~K Baccus}.} \bibinfo{year}{1977}\natexlab{}.
\newblock \showarticletitle{Effect of reflection of feeling and probe on client self-referenced affect.}
\newblock \bibinfo{journal}{\emph{Journal of Counseling Psychology}} \bibinfo{volume}{24}, \bibinfo{number}{5} (\bibinfo{year}{1977}), \bibinfo{pages}{440}.
\newblock


\bibitem[Jain et~al\mbox{.}(2018)]%
        {jain2018evaluating}
\bibfield{author}{\bibinfo{person}{Mohit Jain}, \bibinfo{person}{Pratyush Kumar}, \bibinfo{person}{Ramachandra Kota}, {and} \bibinfo{person}{Shwetak~N Patel}.} \bibinfo{year}{2018}\natexlab{}.
\newblock \showarticletitle{Evaluating and informing the design of chatbots}. In \bibinfo{booktitle}{\emph{Proceedings of the 2018 designing interactive systems conference}}. \bibinfo{pages}{895--906}.
\newblock


\bibitem[Jo et~al\mbox{.}(2024)]%
        {jo2024understanding}
\bibfield{author}{\bibinfo{person}{Eunkyung Jo}, \bibinfo{person}{Yuin Jeong}, \bibinfo{person}{SoHyun Park}, \bibinfo{person}{Daniel~A Epstein}, {and} \bibinfo{person}{Young-Ho Kim}.} \bibinfo{year}{2024}\natexlab{}.
\newblock \showarticletitle{Understanding the Impact of Long-Term Memory on Self-Disclosure with Large Language Model-Driven Chatbots for Public Health Intervention}.
\newblock \bibinfo{journal}{\emph{arXiv preprint arXiv:2402.11353}} (\bibinfo{year}{2024}).
\newblock


\bibitem[Joyce et~al\mbox{.}(2016)]%
        {joyce2016mobile}
\bibfield{author}{\bibinfo{person}{Ger Joyce}, \bibinfo{person}{Mariana Lilley}, \bibinfo{person}{Trevor Barker}, {and} \bibinfo{person}{Amanda Jefferies}.} \bibinfo{year}{2016}\natexlab{}.
\newblock \showarticletitle{Mobile application tutorials: perception of usefulness from an HCI expert perspective}. In \bibinfo{booktitle}{\emph{Human-Computer Interaction. Interaction Platforms and Techniques: 18th International Conference, HCI International 2016, Toronto, ON, Canada, July 17-22, 2016. Proceedings, Part II 18}}. Springer, \bibinfo{pages}{302--308}.
\newblock


\bibitem[Ke et~al\mbox{.}(2024)]%
        {ke2024development}
\bibfield{author}{\bibinfo{person}{YuHe Ke}, \bibinfo{person}{Liyuan Jin}, \bibinfo{person}{Kabilan Elangovan}, \bibinfo{person}{Hairil~Rizal Abdullah}, \bibinfo{person}{Nan Liu}, \bibinfo{person}{Alex Tiong~Heng Sia}, \bibinfo{person}{Chai~Rick Soh}, \bibinfo{person}{Joshua Yi~Min Tung}, \bibinfo{person}{Jasmine Chiat~Ling Ong}, {and} \bibinfo{person}{Daniel Shu~Wei Ting}.} \bibinfo{year}{2024}\natexlab{}.
\newblock \showarticletitle{Development and Testing of Retrieval Augmented Generation in Large Language Models--A Case Study Report}.
\newblock \bibinfo{journal}{\emph{arXiv preprint arXiv:2402.01733}} (\bibinfo{year}{2024}).
\newblock


\bibitem[Kef and Dekovi{\'c}(2004)]%
        {kef2004role}
\bibfield{author}{\bibinfo{person}{Sabina Kef} {and} \bibinfo{person}{Maja Dekovi{\'c}}.} \bibinfo{year}{2004}\natexlab{}.
\newblock \showarticletitle{The role of parental and peer support in adolescents well-being: a comparison of adolescents with and without a visual impairment}.
\newblock \bibinfo{journal}{\emph{Journal of adolescence}} \bibinfo{volume}{27}, \bibinfo{number}{4} (\bibinfo{year}{2004}), \bibinfo{pages}{453--466}.
\newblock


\bibitem[Keselman et~al\mbox{.}(1998)]%
        {keselman1998statistical}
\bibfield{author}{\bibinfo{person}{Harvey~J Keselman}, \bibinfo{person}{Carl~J Huberty}, \bibinfo{person}{Lisa~M Lix}, \bibinfo{person}{Stephen Olejnik}, \bibinfo{person}{Robert~A Cribbie}, \bibinfo{person}{Barbara Donahue}, \bibinfo{person}{Rhonda~K Kowalchuk}, \bibinfo{person}{Laureen~L Lowman}, \bibinfo{person}{Martha~D Petoskey}, \bibinfo{person}{Joanne~C Keselman}, {et~al\mbox{.}}} \bibinfo{year}{1998}\natexlab{}.
\newblock \showarticletitle{Statistical practices of educational researchers: An analysis of their ANOVA, MANOVA, and ANCOVA analyses}.
\newblock \bibinfo{journal}{\emph{Review of educational research}} \bibinfo{volume}{68}, \bibinfo{number}{3} (\bibinfo{year}{1998}), \bibinfo{pages}{350--386}.
\newblock


\bibitem[Kharitonova et~al\mbox{.}(2024)]%
        {kharitonova2024incorporating}
\bibfield{author}{\bibinfo{person}{Ksenia Kharitonova}, \bibinfo{person}{David P{\'e}rez-Fern{\'a}ndez}, \bibinfo{person}{Javier Guti{\'e}rrez-Hernando}, \bibinfo{person}{Asier Guti{\'e}rrez-Fandi{\~n}o}, \bibinfo{person}{Zoraida Callejas}, {and} \bibinfo{person}{David Griol}.} \bibinfo{year}{2024}\natexlab{}.
\newblock \showarticletitle{Incorporating evidence into mental health Q\&A: a novel method to use generative language models for validated clinical content extraction}.
\newblock \bibinfo{journal}{\emph{Behaviour \& Information Technology}} (\bibinfo{year}{2024}), \bibinfo{pages}{1--18}.
\newblock


\bibitem[Koppula and Saxena(2015)]%
        {koppula2015anticipating}
\bibfield{author}{\bibinfo{person}{Hema~S Koppula} {and} \bibinfo{person}{Ashutosh Saxena}.} \bibinfo{year}{2015}\natexlab{}.
\newblock \showarticletitle{Anticipating human activities using object affordances for reactive robotic response}.
\newblock \bibinfo{journal}{\emph{IEEE transactions on pattern analysis and machine intelligence}} \bibinfo{volume}{38}, \bibinfo{number}{1} (\bibinfo{year}{2015}), \bibinfo{pages}{14--29}.
\newblock


\bibitem[Kornfield et~al\mbox{.}(2022)]%
        {kornfield2022meeting}
\bibfield{author}{\bibinfo{person}{Rachel Kornfield}, \bibinfo{person}{Jonah Meyerhoff}, \bibinfo{person}{Hannah Studd}, \bibinfo{person}{Ananya Bhattacharjee}, \bibinfo{person}{Joseph~Jay Williams}, \bibinfo{person}{Madhu Reddy}, {and} \bibinfo{person}{David~C Mohr}.} \bibinfo{year}{2022}\natexlab{}.
\newblock \showarticletitle{Meeting users where they are: user-centered design of an automated text messaging tool to support the mental health of young adults}. In \bibinfo{booktitle}{\emph{Proceedings of the 2022 CHI Conference on Human Factors in Computing Systems}}. \bibinfo{pages}{1--16}.
\newblock


\bibitem[Lee et~al\mbox{.}(2021)]%
        {lee2021restatement}
\bibfield{author}{\bibinfo{person}{John~SY Lee}, \bibinfo{person}{Baikun Liang}, {and} \bibinfo{person}{Haley~HM Fong}.} \bibinfo{year}{2021}\natexlab{}.
\newblock \showarticletitle{Restatement and question generation for counsellor chatbot}. In \bibinfo{booktitle}{\emph{1st Workshop on Natural Language Processing for Programming (NLP4Prog)}}. Association for Computational Linguistics (ACL), \bibinfo{pages}{1--7}.
\newblock


\bibitem[Lee et~al\mbox{.}(2019)]%
        {lee2019caring}
\bibfield{author}{\bibinfo{person}{Minha Lee}, \bibinfo{person}{Sander Ackermans}, \bibinfo{person}{Nena Van~As}, \bibinfo{person}{Hanwen Chang}, \bibinfo{person}{Enzo Lucas}, {and} \bibinfo{person}{Wijnand IJsselsteijn}.} \bibinfo{year}{2019}\natexlab{}.
\newblock \showarticletitle{Caring for Vincent: a chatbot for self-compassion}. In \bibinfo{booktitle}{\emph{Proceedings of the 2019 CHI Conference on Human Factors in Computing Systems}}. \bibinfo{pages}{1--13}.
\newblock


\bibitem[Lee et~al\mbox{.}(2020a)]%
        {lee2020designing}
\bibfield{author}{\bibinfo{person}{Yi-Chieh Lee}, \bibinfo{person}{Naomi Yamashita}, {and} \bibinfo{person}{Yun Huang}.} \bibinfo{year}{2020}\natexlab{a}.
\newblock \showarticletitle{Designing a chatbot as a mediator for promoting deep self-disclosure to a real mental health professional}.
\newblock \bibinfo{journal}{\emph{Proceedings of the ACM on Human-Computer Interaction}} \bibinfo{volume}{4}, \bibinfo{number}{CSCW1} (\bibinfo{year}{2020}), \bibinfo{pages}{1--27}.
\newblock


\bibitem[Lee et~al\mbox{.}(2020b)]%
        {lee2020hear}
\bibfield{author}{\bibinfo{person}{Yi-Chieh Lee}, \bibinfo{person}{Naomi Yamashita}, \bibinfo{person}{Yun Huang}, {and} \bibinfo{person}{Wai Fu}.} \bibinfo{year}{2020}\natexlab{b}.
\newblock \showarticletitle{" I hear you, I feel you": encouraging deep self-disclosure through a chatbot}. In \bibinfo{booktitle}{\emph{Proceedings of the 2020 CHI conference on human factors in computing systems}}. \bibinfo{pages}{1--12}.
\newblock


\bibitem[Li et~al\mbox{.}(2023b)]%
        {li2023understanding}
\bibfield{author}{\bibinfo{person}{Anqi Li}, \bibinfo{person}{Lizhi Ma}, \bibinfo{person}{Yaling Mei}, \bibinfo{person}{Hongliang He}, \bibinfo{person}{Shuai Zhang}, \bibinfo{person}{Huachuan Qiu}, {and} \bibinfo{person}{Zhenzhong Lan}.} \bibinfo{year}{2023}\natexlab{b}.
\newblock \showarticletitle{Understanding client reactions in online mental health counseling}.
\newblock \bibinfo{journal}{\emph{arXiv preprint arXiv:2306.15334}} (\bibinfo{year}{2023}).
\newblock


\bibitem[Li et~al\mbox{.}(2023a)]%
        {li2023chatharuhi}
\bibfield{author}{\bibinfo{person}{Cheng Li}, \bibinfo{person}{Ziang Leng}, \bibinfo{person}{Chenxi Yan}, \bibinfo{person}{Junyi Shen}, \bibinfo{person}{Hao Wang}, \bibinfo{person}{Weishi Mi}, \bibinfo{person}{Yaying Fei}, \bibinfo{person}{Xiaoyang Feng}, \bibinfo{person}{Song Yan}, \bibinfo{person}{HaoSheng Wang}, {et~al\mbox{.}}} \bibinfo{year}{2023}\natexlab{a}.
\newblock \showarticletitle{Chatharuhi: Reviving anime character in reality via large language model}.
\newblock \bibinfo{journal}{\emph{arXiv preprint arXiv:2308.09597}} (\bibinfo{year}{2023}).
\newblock


\bibitem[Li et~al\mbox{.}(2024)]%
        {Li_Wu_Liu_Zhang_Guo_Peng_2024}
\bibfield{author}{\bibinfo{person}{Shuailin Li}, \bibinfo{person}{Shiwei Wu}, \bibinfo{person}{Tianjian Liu}, \bibinfo{person}{Han Zhang}, \bibinfo{person}{Qingyu Guo}, {and} \bibinfo{person}{Zhenhui Peng}.} \bibinfo{year}{2024}\natexlab{}.
\newblock \showarticletitle{Understanding the Features of Text-Image Posts and Their Received Social Support in Online Grief Support Communities}.
\newblock \bibinfo{journal}{\emph{Proceedings of the International AAAI Conference on Web and Social Media}} \bibinfo{volume}{18}, \bibinfo{number}{1} (\bibinfo{date}{May} \bibinfo{year}{2024}), \bibinfo{pages}{917--929}.
\newblock
\urldef\tempurl%
\url{https://doi.org/10.1609/icwsm.v18i1.31362}
\showDOI{\tempurl}


\bibitem[Liao et~al\mbox{.}(2016)]%
        {liao2016can}
\bibfield{author}{\bibinfo{person}{Q~Vera Liao}, \bibinfo{person}{Matthew Davis}, \bibinfo{person}{Werner Geyer}, \bibinfo{person}{Michael Muller}, {and} \bibinfo{person}{N~Sadat Shami}.} \bibinfo{year}{2016}\natexlab{}.
\newblock \showarticletitle{What can you do? Studying social-agent orientation and agent proactive interactions with an agent for employees}. In \bibinfo{booktitle}{\emph{Proceedings of the 2016 acm conference on designing interactive systems}}. \bibinfo{pages}{264--275}.
\newblock


\bibitem[Liu et~al\mbox{.}(2021a)]%
        {liu-etal-2021-towards}
\bibfield{author}{\bibinfo{person}{Siyang Liu}, \bibinfo{person}{Chujie Zheng}, \bibinfo{person}{Orianna Demasi}, \bibinfo{person}{Sahand Sabour}, \bibinfo{person}{Yu Li}, \bibinfo{person}{Zhou Yu}, \bibinfo{person}{Yong Jiang}, {and} \bibinfo{person}{Minlie Huang}.} \bibinfo{year}{2021}\natexlab{a}.
\newblock \showarticletitle{Towards Emotional Support Dialog Systems}. In \bibinfo{booktitle}{\emph{Proceedings of the 59th Annual Meeting of the Association for Computational Linguistics and the 11th International Joint Conference on Natural Language Processing (Volume 1: Long Papers)}}, \bibfield{editor}{\bibinfo{person}{Chengqing Zong}, \bibinfo{person}{Fei Xia}, \bibinfo{person}{Wenjie Li}, {and} \bibinfo{person}{Roberto Navigli}} (Eds.). \bibinfo{publisher}{Association for Computational Linguistics}, \bibinfo{address}{Online}, \bibinfo{pages}{3469--3483}.
\newblock
\urldef\tempurl%
\url{https://doi.org/10.18653/v1/2021.acl-long.269}
\showDOI{\tempurl}


\bibitem[Liu et~al\mbox{.}(2021b)]%
        {liu2021towards}
\bibfield{author}{\bibinfo{person}{Siyang Liu}, \bibinfo{person}{Chujie Zheng}, \bibinfo{person}{Orianna Demasi}, \bibinfo{person}{Sahand Sabour}, \bibinfo{person}{Yu Li}, \bibinfo{person}{Zhou Yu}, \bibinfo{person}{Yong Jiang}, {and} \bibinfo{person}{Minlie Huang}.} \bibinfo{year}{2021}\natexlab{b}.
\newblock \showarticletitle{Towards emotional support dialog systems}.
\newblock \bibinfo{journal}{\emph{arXiv preprint arXiv:2106.01144}} (\bibinfo{year}{2021}).
\newblock


\bibitem[Lohse et~al\mbox{.}(2014)]%
        {lohse2014robot}
\bibfield{author}{\bibinfo{person}{Manja Lohse}, \bibinfo{person}{Reinier Rothuis}, \bibinfo{person}{Jorge Gallego-P{\'e}rez}, \bibinfo{person}{Daphne~E Karreman}, {and} \bibinfo{person}{Vanessa Evers}.} \bibinfo{year}{2014}\natexlab{}.
\newblock \showarticletitle{Robot gestures make difficult tasks easier: the impact of gestures on perceived workload and task performance}. In \bibinfo{booktitle}{\emph{Proceedings of the SIGCHI conference on human factors in computing systems}}. \bibinfo{pages}{1459--1466}.
\newblock


\bibitem[McConchie et~al\mbox{.}(2022)]%
        {mcconchie2022little}
\bibfield{author}{\bibinfo{person}{James McConchie}, \bibinfo{person}{Brittany~J Hite}, \bibinfo{person}{M~Betsy Blackard}, {and} \bibinfo{person}{Ryan Cheuk~Ming Cheung}.} \bibinfo{year}{2022}\natexlab{}.
\newblock \showarticletitle{With a little help from my friends: Development and validation of the positive peer influence inventory}.
\newblock \bibinfo{journal}{\emph{Applied Developmental Science}} \bibinfo{volume}{26}, \bibinfo{number}{1} (\bibinfo{year}{2022}), \bibinfo{pages}{74--93}.
\newblock


\bibitem[Mead et~al\mbox{.}(2001)]%
        {mead2001peer}
\bibfield{author}{\bibinfo{person}{Shery Mead}, \bibinfo{person}{David Hilton}, {and} \bibinfo{person}{Laurie Curtis}.} \bibinfo{year}{2001}\natexlab{}.
\newblock \showarticletitle{Peer support: a theoretical perspective.}
\newblock \bibinfo{journal}{\emph{Psychiatric rehabilitation journal}} \bibinfo{volume}{25}, \bibinfo{number}{2} (\bibinfo{year}{2001}), \bibinfo{pages}{134}.
\newblock


\bibitem[Medhi~Thies et~al\mbox{.}(2017)]%
        {medhi2017you}
\bibfield{author}{\bibinfo{person}{Indrani Medhi~Thies}, \bibinfo{person}{Nandita Menon}, \bibinfo{person}{Sneha Magapu}, \bibinfo{person}{Manisha Subramony}, {and} \bibinfo{person}{Jacki O’neill}.} \bibinfo{year}{2017}\natexlab{}.
\newblock \showarticletitle{How do you want your chatbot? An exploratory Wizard-of-Oz study with young, urban Indians}. In \bibinfo{booktitle}{\emph{Human-Computer Interaction-INTERACT 2017: 16th IFIP TC 13 International Conference, Mumbai, India, September 25--29, 2017, Proceedings, Part I 16}}. Springer, \bibinfo{pages}{441--459}.
\newblock


\bibitem[Meurisch et~al\mbox{.}(2020)]%
        {meurisch2020exploring}
\bibfield{author}{\bibinfo{person}{Christian Meurisch}, \bibinfo{person}{Cristina~A Mihale-Wilson}, \bibinfo{person}{Adrian Hawlitschek}, \bibinfo{person}{Florian Giger}, \bibinfo{person}{Florian M{\"u}ller}, \bibinfo{person}{Oliver Hinz}, {and} \bibinfo{person}{Max M{\"u}hlh{\"a}user}.} \bibinfo{year}{2020}\natexlab{}.
\newblock \showarticletitle{Exploring user expectations of proactive AI systems}.
\newblock \bibinfo{journal}{\emph{Proceedings of the ACM on Interactive, Mobile, Wearable and Ubiquitous Technologies}} \bibinfo{volume}{4}, \bibinfo{number}{4} (\bibinfo{year}{2020}), \bibinfo{pages}{1--22}.
\newblock


\bibitem[Mou and Xu(2017)]%
        {mou2017media}
\bibfield{author}{\bibinfo{person}{Yi Mou} {and} \bibinfo{person}{Kun Xu}.} \bibinfo{year}{2017}\natexlab{}.
\newblock \showarticletitle{The media inequality: Comparing the initial human-human and human-AI social interactions}.
\newblock \bibinfo{journal}{\emph{Computers in Human Behavior}}  \bibinfo{volume}{72} (\bibinfo{year}{2017}), \bibinfo{pages}{432--440}.
\newblock


\bibitem[Narain et~al\mbox{.}(2020)]%
        {narain2020promoting}
\bibfield{author}{\bibinfo{person}{Jaya Narain}, \bibinfo{person}{Tina Quach}, \bibinfo{person}{Monique Davey}, \bibinfo{person}{Hae~Won Park}, \bibinfo{person}{Cynthia Breazeal}, {and} \bibinfo{person}{Rosalind Picard}.} \bibinfo{year}{2020}\natexlab{}.
\newblock \showarticletitle{Promoting wellbeing with Sunny, a chatbot that facilitates positive messages within social groups}. In \bibinfo{booktitle}{\emph{Extended abstracts of the 2020 CHI conference on human factors in computing systems}}. \bibinfo{pages}{1--8}.
\newblock


\bibitem[Nordberg et~al\mbox{.}(2019)]%
        {nordberg2019designing}
\bibfield{author}{\bibinfo{person}{Oda~Elise Nordberg}, \bibinfo{person}{Jo~Dugstad Wake}, \bibinfo{person}{Emilie~Sektnan Nordby}, \bibinfo{person}{Eivind Flobak}, \bibinfo{person}{Tine Nordgreen}, \bibinfo{person}{Suresh~Kumar Mukhiya}, {and} \bibinfo{person}{Frode Guribye}.} \bibinfo{year}{2019}\natexlab{}.
\newblock \showarticletitle{Designing chatbots for guiding online peer support conversations for adults with ADHD}. In \bibinfo{booktitle}{\emph{International workshop on Chatbot research and design}}. Springer, \bibinfo{pages}{113--126}.
\newblock


\bibitem[O'Dea and Campbell(2011)]%
        {o2011healthy}
\bibfield{author}{\bibinfo{person}{Bridianne O'Dea} {and} \bibinfo{person}{Andrew Campbell}.} \bibinfo{year}{2011}\natexlab{}.
\newblock \showarticletitle{Healthy connections: online social networks and their potential for peer support}.
\newblock In \bibinfo{booktitle}{\emph{Health Informatics: The Transformative Power of Innovation}}. \bibinfo{publisher}{IOS Press}, \bibinfo{pages}{133--140}.
\newblock


\bibitem[O'Leary et~al\mbox{.}(2017)]%
        {o2017design}
\bibfield{author}{\bibinfo{person}{Kathleen O'Leary}, \bibinfo{person}{Arpita Bhattacharya}, \bibinfo{person}{Sean~A Munson}, \bibinfo{person}{Jacob~O Wobbrock}, {and} \bibinfo{person}{Wanda Pratt}.} \bibinfo{year}{2017}\natexlab{}.
\newblock \showarticletitle{Design opportunities for mental health peer support technologies}. In \bibinfo{booktitle}{\emph{Proceedings of the 2017 ACM conference on computer supported cooperative work and social computing}}. \bibinfo{pages}{1470--1484}.
\newblock


\bibitem[O'Leary et~al\mbox{.}(2018)]%
        {o2018suddenly}
\bibfield{author}{\bibinfo{person}{Kathleen O'Leary}, \bibinfo{person}{Stephen~M Schueller}, \bibinfo{person}{Jacob~O Wobbrock}, {and} \bibinfo{person}{Wanda Pratt}.} \bibinfo{year}{2018}\natexlab{}.
\newblock \showarticletitle{“Suddenly, we got to become therapists for each other” Designing Peer Support Chats for Mental Health}. In \bibinfo{booktitle}{\emph{Proceedings of the 2018 CHI conference on human factors in computing systems}}. \bibinfo{pages}{1--14}.
\newblock


\bibitem[Pan et~al\mbox{.}(2023)]%
        {pan2023desirable}
\bibfield{author}{\bibinfo{person}{Shuyi Pan}, \bibinfo{person}{Jie Cui}, {and} \bibinfo{person}{Yi Mou}.} \bibinfo{year}{2023}\natexlab{}.
\newblock \showarticletitle{Desirable or Distasteful? Exploring Uncertainty in Human-Chatbot Relationships}.
\newblock \bibinfo{journal}{\emph{International Journal of Human--Computer Interaction}} (\bibinfo{year}{2023}), \bibinfo{pages}{1--11}.
\newblock


\bibitem[Park et~al\mbox{.}(2023)]%
        {park2023generative}
\bibfield{author}{\bibinfo{person}{Joon~Sung Park}, \bibinfo{person}{Joseph O'Brien}, \bibinfo{person}{Carrie~Jun Cai}, \bibinfo{person}{Meredith~Ringel Morris}, \bibinfo{person}{Percy Liang}, {and} \bibinfo{person}{Michael~S Bernstein}.} \bibinfo{year}{2023}\natexlab{}.
\newblock \showarticletitle{Generative agents: Interactive simulacra of human behavior}. In \bibinfo{booktitle}{\emph{Proceedings of the 36th Annual ACM Symposium on User Interface Software and Technology}}. \bibinfo{pages}{1--22}.
\newblock


\bibitem[Pascoe et~al\mbox{.}(2020)]%
        {pascoe2020impact}
\bibfield{author}{\bibinfo{person}{Michaela~C Pascoe}, \bibinfo{person}{Sarah~E Hetrick}, {and} \bibinfo{person}{Alexandra~G Parker}.} \bibinfo{year}{2020}\natexlab{}.
\newblock \showarticletitle{The impact of stress on students in secondary school and higher education}.
\newblock \bibinfo{journal}{\emph{International journal of adolescence and youth}} \bibinfo{volume}{25}, \bibinfo{number}{1} (\bibinfo{year}{2020}), \bibinfo{pages}{104--112}.
\newblock


\bibitem[Peng et~al\mbox{.}(2020)]%
        {peng2020exploring}
\bibfield{author}{\bibinfo{person}{Zhenhui Peng}, \bibinfo{person}{Qingyu Guo}, \bibinfo{person}{Ka~Wing Tsang}, {and} \bibinfo{person}{Xiaojuan Ma}.} \bibinfo{year}{2020}\natexlab{}.
\newblock \showarticletitle{Exploring the effects of technological writing assistance for support providers in online mental health community}. In \bibinfo{booktitle}{\emph{Proceedings of the 2020 CHI Conference on Human Factors in Computing Systems}}. \bibinfo{pages}{1--15}.
\newblock


\bibitem[Peng et~al\mbox{.}(2019)]%
        {peng2019design}
\bibfield{author}{\bibinfo{person}{Zhenhui Peng}, \bibinfo{person}{Yunhwan Kwon}, \bibinfo{person}{Jiaan Lu}, \bibinfo{person}{Ziming Wu}, {and} \bibinfo{person}{Xiaojuan Ma}.} \bibinfo{year}{2019}\natexlab{}.
\newblock \showarticletitle{Design and evaluation of service robot's proactivity in decision-making support process}. In \bibinfo{booktitle}{\emph{proceedings of the 2019 CHI conference on human factors in computing systems}}. \bibinfo{pages}{1--13}.
\newblock


\bibitem[Peng et~al\mbox{.}(2021)]%
        {peng_chi21}
\bibfield{author}{\bibinfo{person}{Zhenhui Peng}, \bibinfo{person}{Xiaojuan Ma}, \bibinfo{person}{Diyi Yang}, \bibinfo{person}{Ka~Wing Tsang}, {and} \bibinfo{person}{Qingyu Guo}.} \bibinfo{year}{2021}\natexlab{}.
\newblock \showarticletitle{Effects of Support-Seekers’ Community Knowledge on Their Expressed Satisfaction with the Received Comments in Mental Health Communities}. In \bibinfo{booktitle}{\emph{Proceedings of the 2021 CHI Conference on Human Factors in Computing Systems}} (Yokohama, Japan) \emph{(\bibinfo{series}{CHI '21})}. \bibinfo{publisher}{Association for Computing Machinery}, \bibinfo{address}{New York, NY, USA}, Article \bibinfo{articleno}{536}, \bibinfo{numpages}{12}~pages.
\newblock
\showISBNx{9781450380966}
\urldef\tempurl%
\url{https://doi.org/10.1145/3411764.3445446}
\showDOI{\tempurl}


\bibitem[Pfeiffer et~al\mbox{.}(2011)]%
        {pfeiffer2011efficacy}
\bibfield{author}{\bibinfo{person}{Paul~N Pfeiffer}, \bibinfo{person}{Michele Heisler}, \bibinfo{person}{John~D Piette}, \bibinfo{person}{Mary~AM Rogers}, {and} \bibinfo{person}{Marcia Valenstein}.} \bibinfo{year}{2011}\natexlab{}.
\newblock \showarticletitle{Efficacy of peer support interventions for depression: a meta-analysis}.
\newblock \bibinfo{journal}{\emph{General hospital psychiatry}} \bibinfo{volume}{33}, \bibinfo{number}{1} (\bibinfo{year}{2011}), \bibinfo{pages}{29--36}.
\newblock


\bibitem[Portela and Granell-Canut(2017)]%
        {portela2017new}
\bibfield{author}{\bibinfo{person}{Manuel Portela} {and} \bibinfo{person}{Carlos Granell-Canut}.} \bibinfo{year}{2017}\natexlab{}.
\newblock \showarticletitle{A new friend in our smartphone? Observing interactions with chatbots in the search of emotional engagement}. In \bibinfo{booktitle}{\emph{Proceedings of the XVIII International Conference on Human Computer Interaction}}. \bibinfo{pages}{1--7}.
\newblock


\bibitem[Potts et~al\mbox{.}(2023)]%
        {potts2023multilingual}
\bibfield{author}{\bibinfo{person}{Courtney Potts}, \bibinfo{person}{Frida Lindstr{\"o}m}, \bibinfo{person}{Raymond Bond}, \bibinfo{person}{Maurice Mulvenna}, \bibinfo{person}{Frederick Booth}, \bibinfo{person}{Edel Ennis}, \bibinfo{person}{Karolina Parding}, \bibinfo{person}{Catrine Kostenius}, \bibinfo{person}{Thomas Broderick}, \bibinfo{person}{Kyle Boyd}, {et~al\mbox{.}}} \bibinfo{year}{2023}\natexlab{}.
\newblock \showarticletitle{A multilingual digital mental health and well-being Chatbot (ChatPal): pre-post multicenter intervention study}.
\newblock \bibinfo{journal}{\emph{Journal of Medical Internet Research}}  \bibinfo{volume}{25} (\bibinfo{year}{2023}), \bibinfo{pages}{e43051}.
\newblock


\bibitem[Qiu et~al\mbox{.}(2023)]%
        {qiu2023psychat}
\bibfield{author}{\bibinfo{person}{Huachuan Qiu}, \bibinfo{person}{Anqi Li}, \bibinfo{person}{Lizhi Ma}, {and} \bibinfo{person}{Zhenzhong Lan}.} \bibinfo{year}{2023}\natexlab{}.
\newblock \showarticletitle{PsyChat: A Client-Centric Dialogue System for Mental Health Support}.
\newblock \bibinfo{journal}{\emph{arXiv preprint arXiv:2312.04262}} (\bibinfo{year}{2023}).
\newblock


\bibitem[Rawte et~al\mbox{.}(2023)]%
        {rawte2023survey}
\bibfield{author}{\bibinfo{person}{Vipula Rawte}, \bibinfo{person}{Amit Sheth}, {and} \bibinfo{person}{Amitava Das}.} \bibinfo{year}{2023}\natexlab{}.
\newblock \showarticletitle{A survey of hallucination in large foundation models}.
\newblock \bibinfo{journal}{\emph{arXiv preprint arXiv:2309.05922}} (\bibinfo{year}{2023}).
\newblock


\bibitem[Rho et~al\mbox{.}(2018)]%
        {rho2018fostering}
\bibfield{author}{\bibinfo{person}{Eugenia Ha~Rim Rho}, \bibinfo{person}{Gloria Mark}, {and} \bibinfo{person}{Melissa Mazmanian}.} \bibinfo{year}{2018}\natexlab{}.
\newblock \showarticletitle{Fostering civil discourse online: Linguistic behavior in comments of\# metoo articles across political perspectives}.
\newblock \bibinfo{journal}{\emph{Proceedings of the ACM on human-computer interaction}} \bibinfo{volume}{2}, \bibinfo{number}{CSCW} (\bibinfo{year}{2018}), \bibinfo{pages}{1--28}.
\newblock


\bibitem[Robotham and Julian(2006)]%
        {robotham2006stress}
\bibfield{author}{\bibinfo{person}{David Robotham} {and} \bibinfo{person}{Claire Julian}.} \bibinfo{year}{2006}\natexlab{}.
\newblock \showarticletitle{Stress and the higher education student: A critical review of the literature}.
\newblock \bibinfo{journal}{\emph{Journal of further and higher education}} \bibinfo{volume}{30}, \bibinfo{number}{02} (\bibinfo{year}{2006}), \bibinfo{pages}{107--117}.
\newblock


\bibitem[Salminen et~al\mbox{.}(2018)]%
        {salminen2018persona}
\bibfield{author}{\bibinfo{person}{Joni Salminen}, \bibinfo{person}{Haewoon Kwak}, \bibinfo{person}{Jo{\~a}o~M Santos}, \bibinfo{person}{Soon-Gyo Jung}, \bibinfo{person}{Jisun An}, {and} \bibinfo{person}{Bernard~J Jansen}.} \bibinfo{year}{2018}\natexlab{}.
\newblock \showarticletitle{Persona perception scale: developing and validating an instrument for human-like representations of data}. In \bibinfo{booktitle}{\emph{Extended Abstracts of the 2018 CHI Conference on Human Factors in Computing Systems}}. \bibinfo{pages}{1--6}.
\newblock


\bibitem[Schwarzer and Knoll(2007)]%
        {schwarzer2007functional}
\bibfield{author}{\bibinfo{person}{Ralf Schwarzer} {and} \bibinfo{person}{Nina Knoll}.} \bibinfo{year}{2007}\natexlab{}.
\newblock \showarticletitle{Functional roles of social support within the stress and coping process: A theoretical and empirical overview}.
\newblock \bibinfo{journal}{\emph{International journal of psychology}} \bibinfo{volume}{42}, \bibinfo{number}{4} (\bibinfo{year}{2007}), \bibinfo{pages}{243--252}.
\newblock


\bibitem[Schwarzer and Leppin(1991)]%
        {schwarzer1991social}
\bibfield{author}{\bibinfo{person}{Ralf Schwarzer} {and} \bibinfo{person}{Anja Leppin}.} \bibinfo{year}{1991}\natexlab{}.
\newblock \showarticletitle{Social support and health: A theoretical and empirical overview}.
\newblock \bibinfo{journal}{\emph{Journal of social and personal relationships}} \bibinfo{volume}{8}, \bibinfo{number}{1} (\bibinfo{year}{1991}), \bibinfo{pages}{99--127}.
\newblock


\bibitem[Shum et~al\mbox{.}(2018)]%
        {shum2018eliza}
\bibfield{author}{\bibinfo{person}{Heung-Yeung Shum}, \bibinfo{person}{Xiao-dong He}, {and} \bibinfo{person}{Di Li}.} \bibinfo{year}{2018}\natexlab{}.
\newblock \showarticletitle{From Eliza to XiaoIce: challenges and opportunities with social chatbots}.
\newblock \bibinfo{journal}{\emph{Frontiers of Information Technology \& Electronic Engineering}}  \bibinfo{volume}{19} (\bibinfo{year}{2018}), \bibinfo{pages}{10--26}.
\newblock


\bibitem[{Significant Gravitas}({[n.\,d.]})]%
        {Significant_Gravitas_AutoGPT}
\bibfield{author}{\bibinfo{person}{{Significant Gravitas}}.} \bibinfo{year}{[n.\,d.]}\natexlab{}.
\newblock \bibinfo{booktitle}{\emph{{AutoGPT}}}.
\newblock
\urldef\tempurl%
\url{https://github.com/Significant-Gravitas/AutoGPT}
\showURL{%
\tempurl}


\bibitem[Silvervarg and J{\"o}nsson(2013)]%
        {silvervarg2013iterative}
\bibfield{author}{\bibinfo{person}{Annika Silvervarg} {and} \bibinfo{person}{Arne J{\"o}nsson}.} \bibinfo{year}{2013}\natexlab{}.
\newblock \showarticletitle{Iterative development and evaluation of a social conversational agent}. In \bibinfo{booktitle}{\emph{Proceedings of the Sixth International Joint Conference on Natural Language Processing}}. \bibinfo{pages}{1223--1229}.
\newblock


\bibitem[Skjuve et~al\mbox{.}(2021)]%
        {skjuve2021my}
\bibfield{author}{\bibinfo{person}{Marita Skjuve}, \bibinfo{person}{Asbj{\o}rn F{\o}lstad}, \bibinfo{person}{Knut~Inge Fostervold}, {and} \bibinfo{person}{Petter~Bae Brandtzaeg}.} \bibinfo{year}{2021}\natexlab{}.
\newblock \showarticletitle{My chatbot companion-a study of human-chatbot relationships}.
\newblock \bibinfo{journal}{\emph{International Journal of Human-Computer Studies}}  \bibinfo{volume}{149} (\bibinfo{year}{2021}), \bibinfo{pages}{102601}.
\newblock


\bibitem[Skjuve et~al\mbox{.}(2022)]%
        {skjuve2022longitudinal}
\bibfield{author}{\bibinfo{person}{Marita Skjuve}, \bibinfo{person}{Asbj{\o}rn F{\o}lstad}, \bibinfo{person}{Knut~Inge Fostervold}, {and} \bibinfo{person}{Petter~Bae Brandtzaeg}.} \bibinfo{year}{2022}\natexlab{}.
\newblock \showarticletitle{A longitudinal study of human--chatbot relationships}.
\newblock \bibinfo{journal}{\emph{International Journal of Human-Computer Studies}}  \bibinfo{volume}{168} (\bibinfo{year}{2022}), \bibinfo{pages}{102903}.
\newblock


\bibitem[Sprecher et~al\mbox{.}(2013)]%
        {sprecher2013effects}
\bibfield{author}{\bibinfo{person}{Susan Sprecher}, \bibinfo{person}{Stanislav Treger}, {and} \bibinfo{person}{Joshua~D Wondra}.} \bibinfo{year}{2013}\natexlab{}.
\newblock \showarticletitle{Effects of self-disclosure role on liking, closeness, and other impressions in get-acquainted interactions}.
\newblock \bibinfo{journal}{\emph{Journal of Social and Personal Relationships}} \bibinfo{volume}{30}, \bibinfo{number}{4} (\bibinfo{year}{2013}), \bibinfo{pages}{497--514}.
\newblock


\bibitem[St et~al\mbox{.}(1989)]%
        {st1989analysis}
\bibfield{author}{\bibinfo{person}{Lars St}, \bibinfo{person}{Svante Wold}, {et~al\mbox{.}}} \bibinfo{year}{1989}\natexlab{}.
\newblock \showarticletitle{Analysis of variance (ANOVA)}.
\newblock \bibinfo{journal}{\emph{Chemometrics and intelligent laboratory systems}} \bibinfo{volume}{6}, \bibinfo{number}{4} (\bibinfo{year}{1989}), \bibinfo{pages}{259--272}.
\newblock


\bibitem[Stiles(1980)]%
        {stiles1980measurement}
\bibfield{author}{\bibinfo{person}{William~B Stiles}.} \bibinfo{year}{1980}\natexlab{}.
\newblock \showarticletitle{Measurement of the impact of psychotherapy sessions.}
\newblock \bibinfo{journal}{\emph{Journal of consulting and clinical psychology}} \bibinfo{volume}{48}, \bibinfo{number}{2} (\bibinfo{year}{1980}), \bibinfo{pages}{176}.
\newblock


\bibitem[Sulaiman et~al\mbox{.}(2022)]%
        {sulaiman2022anxiety}
\bibfield{author}{\bibinfo{person}{Suliana Sulaiman}, \bibinfo{person}{Marzita Mansor}, \bibinfo{person}{Rohaizah~Abdul Wahid}, {and} \bibinfo{person}{Nur Anis Alisa~Nor Azhar}.} \bibinfo{year}{2022}\natexlab{}.
\newblock \showarticletitle{Anxiety assistance mobile apps chatbot using cognitive behavioural therapy}.
\newblock \bibinfo{journal}{\emph{International Journal of Artificial Intelligence}} \bibinfo{volume}{9}, \bibinfo{number}{1} (\bibinfo{year}{2022}), \bibinfo{pages}{17--23}.
\newblock


\bibitem[Sumers et~al\mbox{.}(2023)]%
        {sumers2023cognitive}
\bibfield{author}{\bibinfo{person}{Theodore~R Sumers}, \bibinfo{person}{Shunyu Yao}, \bibinfo{person}{Karthik Narasimhan}, {and} \bibinfo{person}{Thomas~L Griffiths}.} \bibinfo{year}{2023}\natexlab{}.
\newblock \showarticletitle{Cognitive architectures for language agents}.
\newblock \bibinfo{journal}{\emph{arXiv preprint arXiv:2309.02427}} (\bibinfo{year}{2023}).
\newblock


\bibitem[Ta et~al\mbox{.}(2020)]%
        {ta2020user}
\bibfield{author}{\bibinfo{person}{Vivian Ta}, \bibinfo{person}{Caroline Griffith}, \bibinfo{person}{Carolynn Boatfield}, \bibinfo{person}{Xinyu Wang}, \bibinfo{person}{Maria Civitello}, \bibinfo{person}{Haley Bader}, \bibinfo{person}{Esther DeCero}, \bibinfo{person}{Alexia Loggarakis}, {et~al\mbox{.}}} \bibinfo{year}{2020}\natexlab{}.
\newblock \showarticletitle{User experiences of social support from companion chatbots in everyday contexts: thematic analysis}.
\newblock \bibinfo{journal}{\emph{Journal of medical Internet research}} \bibinfo{volume}{22}, \bibinfo{number}{3} (\bibinfo{year}{2020}), \bibinfo{pages}{e16235}.
\newblock


\bibitem[Talebirad and Nadiri(2023)]%
        {talebirad2023multi}
\bibfield{author}{\bibinfo{person}{Yashar Talebirad} {and} \bibinfo{person}{Amirhossein Nadiri}.} \bibinfo{year}{2023}\natexlab{}.
\newblock \showarticletitle{Multi-agent collaboration: Harnessing the power of intelligent llm agents}.
\newblock \bibinfo{journal}{\emph{arXiv preprint arXiv:2306.03314}} (\bibinfo{year}{2023}).
\newblock


\bibitem[Tallyn et~al\mbox{.}(2018)]%
        {tallyn2018ethnobot}
\bibfield{author}{\bibinfo{person}{Ella Tallyn}, \bibinfo{person}{Hector Fried}, \bibinfo{person}{Rory Gianni}, \bibinfo{person}{Amy Isard}, {and} \bibinfo{person}{Chris Speed}.} \bibinfo{year}{2018}\natexlab{}.
\newblock \showarticletitle{The ethnobot: Gathering ethnographies in the age of IoT}. In \bibinfo{booktitle}{\emph{Proceedings of the 2018 CHI conference on human factors in computing systems}}. \bibinfo{pages}{1--13}.
\newblock


\bibitem[Tennant(2002)]%
        {tennant2002life}
\bibfield{author}{\bibinfo{person}{Christopher Tennant}.} \bibinfo{year}{2002}\natexlab{}.
\newblock \showarticletitle{Life events, stress and depression: a review of recent findings}.
\newblock \bibinfo{journal}{\emph{Australian \& New Zealand Journal of Psychiatry}} \bibinfo{volume}{36}, \bibinfo{number}{2} (\bibinfo{year}{2002}), \bibinfo{pages}{173--182}.
\newblock


\bibitem[Thoits(1985)]%
        {thoits1985social}
\bibfield{author}{\bibinfo{person}{Peggy~A Thoits}.} \bibinfo{year}{1985}\natexlab{}.
\newblock \showarticletitle{Social support and psychological well-being: Theoretical possibilities}.
\newblock In \bibinfo{booktitle}{\emph{Social support: Theory, research and applications}}. \bibinfo{publisher}{Springer}, \bibinfo{pages}{51--72}.
\newblock


\bibitem[Thoits(2011)]%
        {thoits2011mechanisms}
\bibfield{author}{\bibinfo{person}{Peggy~A Thoits}.} \bibinfo{year}{2011}\natexlab{}.
\newblock \showarticletitle{Mechanisms linking social ties and support to physical and mental health}.
\newblock \bibinfo{journal}{\emph{Journal of health and social behavior}} \bibinfo{volume}{52}, \bibinfo{number}{2} (\bibinfo{year}{2011}), \bibinfo{pages}{145--161}.
\newblock


\bibitem[Toxtli et~al\mbox{.}(2018)]%
        {toxtli2018understanding}
\bibfield{author}{\bibinfo{person}{Carlos Toxtli}, \bibinfo{person}{Andr{\'e}s Monroy-Hern{\'a}ndez}, {and} \bibinfo{person}{Justin Cranshaw}.} \bibinfo{year}{2018}\natexlab{}.
\newblock \showarticletitle{Understanding chatbot-mediated task management}. In \bibinfo{booktitle}{\emph{Proceedings of the 2018 CHI conference on human factors in computing systems}}. \bibinfo{pages}{1--6}.
\newblock


\bibitem[Vogel et~al\mbox{.}(2014)]%
        {vogel2014social}
\bibfield{author}{\bibinfo{person}{Erin~A Vogel}, \bibinfo{person}{Jason~P Rose}, \bibinfo{person}{Lindsay~R Roberts}, {and} \bibinfo{person}{Katheryn Eckles}.} \bibinfo{year}{2014}\natexlab{}.
\newblock \showarticletitle{Social comparison, social media, and self-esteem.}
\newblock \bibinfo{journal}{\emph{Psychology of popular media culture}} \bibinfo{volume}{3}, \bibinfo{number}{4} (\bibinfo{year}{2014}), \bibinfo{pages}{206}.
\newblock


\bibitem[Wang et~al\mbox{.}(2021)]%
        {wang2021cass}
\bibfield{author}{\bibinfo{person}{Liuping Wang}, \bibinfo{person}{Dakuo Wang}, \bibinfo{person}{Feng Tian}, \bibinfo{person}{Zhenhui Peng}, \bibinfo{person}{Xiangmin Fan}, \bibinfo{person}{Zhan Zhang}, \bibinfo{person}{Mo Yu}, \bibinfo{person}{Xiaojuan Ma}, {and} \bibinfo{person}{Hongan Wang}.} \bibinfo{year}{2021}\natexlab{}.
\newblock \showarticletitle{Cass: Towards building a social-support chatbot for online health community}.
\newblock \bibinfo{journal}{\emph{Proceedings of the ACM on Human-Computer Interaction}} \bibinfo{volume}{5}, \bibinfo{number}{CSCW1} (\bibinfo{year}{2021}), \bibinfo{pages}{1--31}.
\newblock


\bibitem[Wang et~al\mbox{.}(2023)]%
        {wang2023humanoid}
\bibfield{author}{\bibinfo{person}{Zhilin Wang}, \bibinfo{person}{Yu~Ying Chiu}, {and} \bibinfo{person}{Yu~Cheung Chiu}.} \bibinfo{year}{2023}\natexlab{}.
\newblock \showarticletitle{Humanoid agents: Platform for simulating human-like generative agents}.
\newblock \bibinfo{journal}{\emph{arXiv preprint arXiv:2310.05418}} (\bibinfo{year}{2023}).
\newblock


\bibitem[Watson(2017)]%
        {watson2017mechanisms}
\bibfield{author}{\bibinfo{person}{Emma Watson}.} \bibinfo{year}{2017}\natexlab{}.
\newblock \showarticletitle{The mechanisms underpinning peer support: a literature review}.
\newblock \bibinfo{journal}{\emph{Journal of Mental Health}} (\bibinfo{year}{2017}).
\newblock


\bibitem[Xiao et~al\mbox{.}(2023)]%
        {xiao2023powering}
\bibfield{author}{\bibinfo{person}{Ziang Xiao}, \bibinfo{person}{Q~Vera Liao}, \bibinfo{person}{Michelle Zhou}, \bibinfo{person}{Tyrone Grandison}, {and} \bibinfo{person}{Yunyao Li}.} \bibinfo{year}{2023}\natexlab{}.
\newblock \showarticletitle{Powering an ai chatbot with expert sourcing to support credible health information access}. In \bibinfo{booktitle}{\emph{Proceedings of the 28th international conference on intelligent user interfaces}}. \bibinfo{pages}{2--18}.
\newblock


\bibitem[Yuan et~al\mbox{.}(2023)]%
        {yuan2023critrainer}
\bibfield{author}{\bibinfo{person}{Kangyu Yuan}, \bibinfo{person}{Hehai Lin}, \bibinfo{person}{Shilei Cao}, \bibinfo{person}{Zhenhui Peng}, \bibinfo{person}{Qingyu Guo}, {and} \bibinfo{person}{Xiaojuan Ma}.} \bibinfo{year}{2023}\natexlab{}.
\newblock \showarticletitle{CriTrainer: An Adaptive Training Tool for Critical Paper Reading}. In \bibinfo{booktitle}{\emph{Proceedings of the 36th Annual ACM Symposium on User Interface Software and Technology}}. \bibinfo{pages}{1--17}.
\newblock


\bibitem[Zhang et~al\mbox{.}(2023)]%
        {zhang2023memory}
\bibfield{author}{\bibinfo{person}{Kai Zhang}, \bibinfo{person}{Fubang Zhao}, \bibinfo{person}{Yangyang Kang}, {and} \bibinfo{person}{Xiaozhong Liu}.} \bibinfo{year}{2023}\natexlab{}.
\newblock \showarticletitle{Memory-augmented llm personalization with short-and long-term memory coordination}.
\newblock \bibinfo{journal}{\emph{arXiv preprint arXiv:2309.11696}} (\bibinfo{year}{2023}).
\newblock


\bibitem[Zhang et~al\mbox{.}(2015a)]%
        {zhang2015human}
\bibfield{author}{\bibinfo{person}{Yu Zhang}, \bibinfo{person}{Vignesh Narayanan}, \bibinfo{person}{Tathagata Chakraborti}, {and} \bibinfo{person}{Subbarao Kambhampati}.} \bibinfo{year}{2015}\natexlab{a}.
\newblock \showarticletitle{A human factors analysis of proactive support in human-robot teaming}. In \bibinfo{booktitle}{\emph{2015 IEEE/RSJ International Conference on Intelligent Robots and Systems (IROS)}}. IEEE, \bibinfo{pages}{3586--3593}.
\newblock


\bibitem[Zhang et~al\mbox{.}(2015b)]%
        {zhang2015capability}
\bibfield{author}{\bibinfo{person}{Yu Zhang}, \bibinfo{person}{Sarath Sreedharan}, {and} \bibinfo{person}{Subbarao Kambhampati}.} \bibinfo{year}{2015}\natexlab{b}.
\newblock \showarticletitle{Capability Models and Their Applications in Planning.}. In \bibinfo{booktitle}{\emph{AAMAS}}. \bibinfo{pages}{1151--1159}.
\newblock


\end{thebibliography}
\end{document}